\documentclass{article} 
\usepackage{arxiv}
\usepackage{amsmath}

\usepackage[utf8]{inputenc} % allow utf-8 input
\usepackage[T1]{fontenc}    % use 8-bit T1 fonts
\usepackage{hyperref}       % hyperlinks
\usepackage{url}            % simple URL typesetting
\usepackage{booktabs}       % professional-quality tables
\usepackage{amsfonts}       % blackboard math symbols
\usepackage{nicefrac}       % compact symbols for 1/2, etc.
\usepackage{microtype}      % microtypography
\usepackage{lipsum}
\usepackage{graphicx}
\graphicspath{ {./images/} }
\usepackage{xcolor}

\usepackage{float}
 
\title{Coupled Ocean-Atmosphere Dynamics in a Machine Learning Earth System Model}
\author{%
    Chenggong Wang \thanks{Correspondence to c.wang@princeton.edu, mpritchard@nvidia.com, jpathak@nvidia.com} \thanks{Work done during an internship at NVIDIA}\\
    Princeton University\\
    \And
    Michael S. Pritchard \footnotemark[1]\\
    NVIDIA \\
    \And
    Noah Brenowitz \\
    NVIDIA \\
    \\
    \And
    Yair Cohen \\
    NVIDIA \\
    \AND
    Boris Bonev \\
    NVIDIA \\
    \And
    Thorsten Kurth \\
    NVIDIA \\
    \And
    Dale Durran \\
    University of Washington\\ NVIDIA \\
    \And
    Jaideep Pathak \footnotemark[1]
    \\
    NVIDIA \\
}

\begin{document}
\maketitle

\begin{abstract}
Seasonal climate forecasts are socioeconomically important for managing the impacts of extreme weather events and for planning in sectors like agriculture and energy. Climate predictability on seasonal timescales is tied to boundary effects of the ocean on the atmosphere and coupled interactions in the ocean-atmosphere system. We present the Ocean-linked-atmosphere (Ola) model, a high-resolution ($0.25^\circ$) Artificial Intelligence/ Machine Learning (AI/ML) coupled earth-system model which separately models the ocean and atmosphere dynamics using an autoregressive Spherical Fourier Neural Operator architecture, with a view towards enabling fast, accurate, large ensemble forecasts on the seasonal timescale. We find that Ola exhibits learned characteristics of ocean-atmosphere coupled dynamics including tropical oceanic waves with appropriate phase speeds, and an internally generated El Niño/Southern Oscillation (ENSO) having realistic amplitude, geographic structure, and vertical structure within the ocean mixed layer. We present initial evidence of skill in forecasting the ENSO which compares favorably to the SPEAR model of the Geophysical Fluid Dynamics Laboratory. 
\end{abstract}

% keywords can be removed

\section*{Significance}
Artificial Intelligence/Machine Learning (AI/ML) emulators of weather forecast models are faster than traditional numerical models by a factor of 10,000 or more. Our work is the first AI/ML emulator that explicitly emulates both the atmosphere and the upper ocean thermal structure. Our coupled ocean-atmosphere Earth System Model (ESM), named Ola, is able to capture characteristic ocean-atmosphere interactions which strongly influence seasonal outlooks of weather. Our work demonstrates a pathway towards accurate forecasts of long-range seasonal climate and uncertainty using AI models. Our model captures the seasonal oscillation known as El Niño/ Southern Oscillation (ENSO) which has a large impact on global weather and is a key source of long-range predictability.  

%%% Suggested section heads:
\section{Introduction} 

Over the past few years, data-driven ML models have demonstrated remarkable predictive skill for medium-range, whole-atmosphere weather forecasting~\cite{bonev2023spherical, bi2023accurate, lam2022graphcast, pathak2022fourcastnet, karlbauer2023advancing, chen2023fengwu, keisler2022forecasting}. Several autoregressive ML models now rival or surpass the performance of the leading physics-based numerical weather prediction (NWP) models used by top meteorological organizations worldwide ~\cite{rasp2023weatherbench,brenowitz2024practical}. This advancement has prompted the European Center for Medium-Range Weather Forecasting (ECMWF) to develop the AIFS~\cite{bouallegue2024aifs}, an experimental AI/ML companion to its Integrated Forecasting System (IFS) model. Once trained, such ML models are orders of magnitude faster and more energy efficient than traditional NWP. They also usefully sidestep the uncertainties intrinsic to subgrid-scale parameterization. 

A next important step for this technology is successful interactive ocean-atmosphere coupling, which is essential for its use beyond medium-range weather prediction to applications of subseasonal-to-seasonal (S2S) forecasting, seasonal forecasting, and climate prediction. Seasonal forecasting, in particular, requires solving a boundary value problem for the atmosphere coupled with an initial value problem for the ocean. Ocean dynamics, which change slowly, provide boundary conditions for the atmosphere~\cite{brankovic1994predictability}. The El Niño / Southern Oscillation (ENSO), characterized by quasi-periodic warming and cooling of the equatorial Pacific, is a leading source of predictability on seasonal timescales, and influences large-scale atmospheric circulation and hydrological cycle changes \cite{walker1924correlations1,mcphaden2006enso}. Successfully predicting ENSO, which relies on ocean-atmosphere interactions and tropical oceanic wave dynamics ~\cite{bjerknes1969atmospheric,cane2005enso}, tests the capability of any global simulation method to capture realistic atmosphere-ocean interactions. Indeed the ability to simulate ENSO in physical climate models was first enabled by early efforts to couple oceanic and atmospheric components in the 1980s\cite{cane1985theory,cane1986experimental}.  

It is a ripe time to explore coupled atmosphere-ocean ML forecasting. Atmosphere emulators have progressed from generating forecasts spanning weeks to those covering many months \cite{weyn2021sub,bonev2023spherical,arcomano2022hybrid} and even decades\cite{watt2023ace,kochkov2023neural}. Meanwhile, there has been significant progress in the creation of ocean emulators on both global ~\cite{xiong2023ai} and regional~\cite{ subel2024building,chattopadhyay2023oceannet} scales. Currently, there are only limited results from fully integrated ML-based coupled atmosphere and ocean forecasting systems -- Cresswell-Clay et al. (in preparation) introduces a parsimonious coupled ocean-atmosphere model that uses a sparse oceanic state vector, including sea surface temperature (SST) and sea surface height (SSH), capable of supporting multi-decadal forecasts. No coupled ML forecasting system has yet attempted to include subsurface ocean state information in its training.  

This paper assesses the feasibility of coupled ML prediction systems with sub-surface ocean state information. We introduce the Ocean-linked-atmosphere (Ola) coupled ocean-atmosphere model that uses the Spherical Fourier Neural Operator (SFNO) architecture to explicitly model ocean dynamics. Its ocean component represents conditions spanning the Eastern Pacific thermocline, from the surface down to 300 meters. These ocean dynamics are conditioned on atmospheric winds, pressure and temperature, and are linked to an SFNO atmosphere model, which is in turn conditioned on sea surface temperature. 

The structure of this paper is organized as follows. In Sec.~\ref{sec:results}, we assess the model's capacity to produce ENSO dynamics by exploring its predictability on seasonal timescales, assessing its geographic structure, and analyzing the dynamic signatures of internally generated tropical oceanic waves. Our objective is to assess whether our model includes coupled ocean-atmosphere interactions that align with the current theoretical understanding of the physical processes driving the onset and demise of ENSO. Sec.~\ref{sec:discussion} concludes with a discussion of the current model's limitations and potential areas for future development. The coupled ML model configuration, training, and inference methodology are presented in Sec. \ref{sec:methods}

\section{Results}\label{sec:results}

All results in this section use the out-of-sample validation set from years 2017-2022.  

\begin{figure}
    \centering
    \includegraphics[scale=0.8]{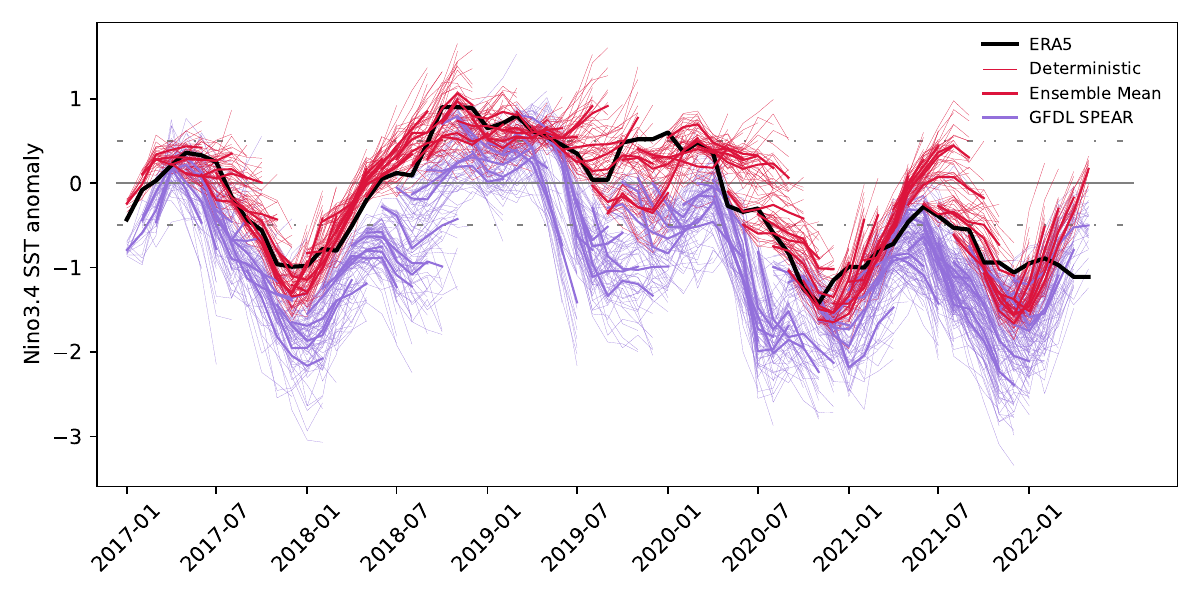}
    \caption{Sea Surface Temperature (SST) anomaly in the Niño 3.4 region of the Pacific Ocean (5$^{\circ}$N to 5$^{\circ}$S, 170$^{\circ}$-120$^{\circ}$W) from 6-month ensemble forecasts generated by the coupled AI/ML ESM Ola(light red lines) and the physics-based model GFDL-SPEAR  (light purple lines). Lagged Ensemble Forecasts (see Sec.~\ref{sec:LEF} for details) are generated at the beginning of every month and run out to 6 months. The ensemble mean for the 12 ensemble forecasts is indicated by corresponding bold lines. The observed ERA5 re-analysis (ground truth) is indicated by the bold black line.  
    }
    \label{fig:enso_plume}
\end{figure}

Our first main finding is that Ola generates realistic amounts of Central Tropical Pacific SST variability in the 5-year validation period (Figure~\ref{fig:enso_plume}). Much like the physics-based baseline--SPEAR model developed at Geophysical Fluid Dynamics Laboratory (GFDL-SPEAR), 6-month Ola forecasts can internally generate both El Niño and La Niña states from a neutral state as well as return to a neutral state when initialized in an El Niño or La Niña state. An exception occurred after the two consecutive La Niña events (winter of 2020/2021 and 2021/2022) when Ola predicted a shift from a La Niña to a neutral state, whereas La Niña persisted in reality. Notably, this triple-dip La Niña was also hard to predict for most operational process-based dynamics and statistics models (see the archived 2021 Nov. forecasts: \url{https://iri.columbia.edu/our-expertise/climate/forecasts/enso/2021-November-quick-look}). It is important to caution that the limited evaluation period  -- necessitated by the need for a sufficiently large training period and limited duration of the UFS Replay dataset -- does not allow us to draw quantitative predictability comparisons between these models. But the finding of qualitatively reasonable SST variability in the Central Pacific is an encouraging initial sign that ML models like Ola may be capable of producing ENSO variability. 
 
Unlike GFDL-SPEAR, Ola produces less of a time-mean cold bias in the Niño 3.4 region (5$^{\circ}$N to 5$^{\circ}$S, 170$^{\circ}$-120$^{\circ}$W). This drift towards systematic equatorial ``cold tongue'' biases is a well-documented issue in many process-based coupled ocean-atmosphere models, including operational seasonal forecast models (e.g. \cite{saha2014ncep, johnson2019seas5, lu2020gfdl}) and climate models \cite{chen2017ensobias,li2014modelbias}. Such biases are linked to the spin-up of the ``double-ITCZ'' rainfall bias in these models, which is influenced by decisions in sub-grid parameterization \cite{yu1999links,song2009convection}. Although postprocessing often tries to mitigate these biases, they tend to degrade ENSO forecasts \cite{Wu2022ColdTongueBias} and lead to associated biases in large-scale atmosphere circulation \cite{wang2020inter} and the hydrological cycle \cite{tian2020double, li2014modelbias}. Interestingly, the absence of tropical drift in Ola suggests that machine learning models may offer new ways to avoid these biases by circumventing subgrid parameterizations, like convection and boundary layer schemes \cite{song2018roles,woelfle2019evolution}, an approach that holds promising potential.

\begin{figure}[t]
    \centering
    \includegraphics[scale=0.6]{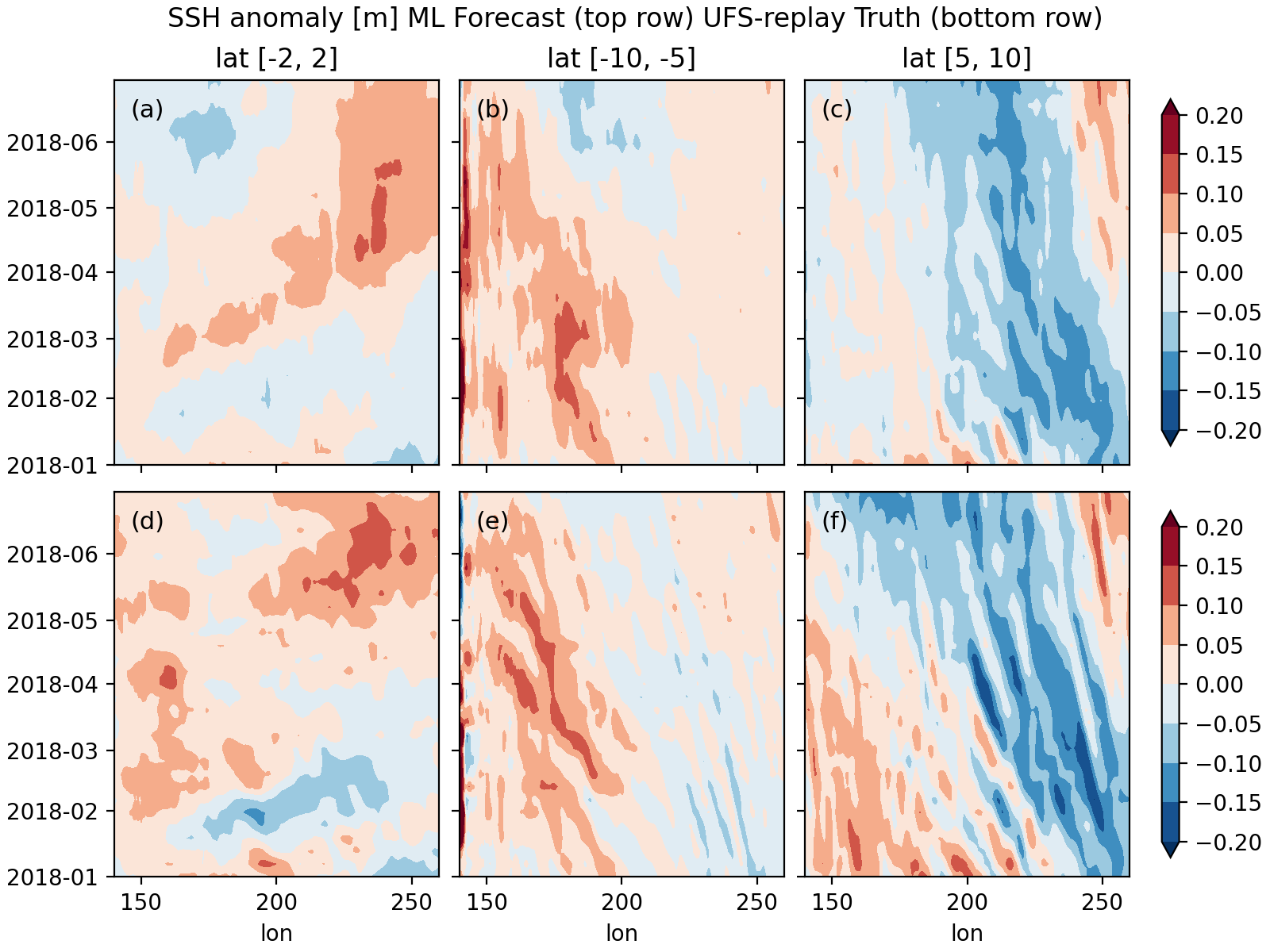}
    \caption{Longitude-time diagrams show the simulations (top row) and corresponding ground truth (UFS-replay dataset) verification (bottom row) of equatorial Kelvin and Rossby waves in the Ola ML ESM initialized on January 1 2018 and simulated out to 6 months. The left column  ((a), (d)) shows the SSH anomaly averaged from 2$^{\circ}$N to 2$^{\circ}$S. The middle column ((b), (e)) and right columns ((c), (f)) shows the SSH anomaly averaged from 5$^{\circ}$ to 10$^{\circ}$ in Southern and Northern Hemisphere. In each panel, the annual mean climatology is subtracted at each grid point to compute the SSH anomaly.  
    }
    \label{fig:eqwave}
\end{figure}

We next ask whether our coupled model generates oceanic equatorial Kelvin and Rossby waves, since their dynamics are critical to ENSO\cite{Sprintall2020,wang2001unified}. This can be assessed qualitatively via the longitude-time evolution of the sea surface height (SSH) anomaly during out-of-sample years. The top row in Figure~\ref{fig:eqwave} shows a single representative example forecast of the SSH from an arbitrary\footnote{The same analysis is repeated across several time periods in the out-of-sample set in Appendix~\ref{sec:app_waves}} initial date using the coupled ML model visualized in three different latitudinal bands straddling and in the vicinity the equator. The corresponding UFS-replay ground truth for the same period is shown in the bottom row of Figure~\ref{fig:eqwave}. Reassuringly, for the latitude band centered on the equator, we observe eastward propagating SSH anomalies having basin crossing timescales of approximately 2 months  --- both initialized and spontaneously generated throughout the 6-month simulation (Figure~\ref{fig:eqwave} (a)). The direction and phase speed of these waves appear to be consistent with observations (Figure~\ref{fig:eqwave} (d)) and the theory of equatorial trapped Kelvin waves for the first baroclinic mode in the ocean \cite{Vallis06}. Off the equator, significantly slower, westward-moving disturbances should be expected, as regulated by Rossby wave dynamics due to the meridional gradient of the Coriolis parameter. Encouragingly, both north (5$^{\circ}$N-10$^{\circ}$N; Figure~\ref{fig:eqwave} (b)) and south ( 5$^{\circ}$S-10$^{\circ}$S; in Figure~\ref{fig:eqwave} (c)) of the equator, westward propagating SSH anomalies are found that propagate with the observed phase speed. Overall, Figure~\ref{fig:eqwave} provides strong evidence that our coupled ML model has learned to predict qualitatively realistic equatorial oceanic wave dynamics, which is a necessary condition to produce realistic ENSO signals.

\begin{figure}
    \centering
    \includegraphics[scale=0.6]{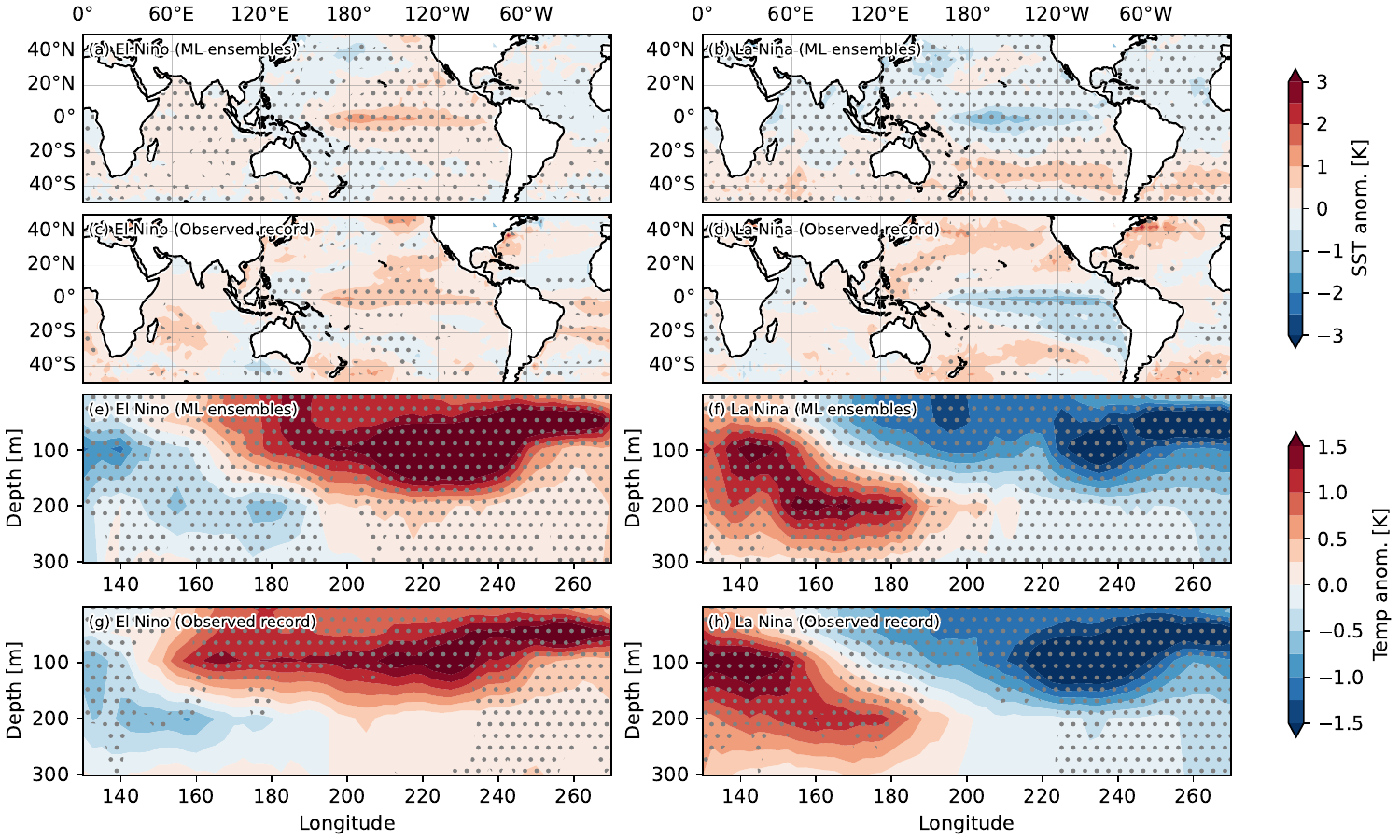}
    \caption{Composite Sea Surface Temperature (a-d) and upper ocean temperature (e-h) anomaly visualizations during El Niño and La Niña events from the Ola coupled ML model ensemble simulations ((a), (b), (e), (f)) at a 4-month lead time and the observed ground truth record ((c), (d), (g), (h)). Ensemble forecasts were generated from the coupled ML model every month from 2017 to 2021 with 12 ensembles at each initialization providing a total of 720 model runs (see Sec. \ref{sec:LEF} for detail). The Ola model forecasts were bias corrected by computing a lead-time dependant model bias from simulations in the training period of 1994-2016 (see Appendix~\ref{sec:app_bias} for bias correction details). The criteria for El Niño and La Niña conditions were defined as the SST monthly anomaly in the Niño 3.4 region exceeding 0.5K. There were 136 months with El Niño states and 187 months with La Niña states at a lead time of 4 months in the 720 ensemble simulations while the observed record during the period from 2017 to 2021 had 13 months with El Niño states and 14 months with La Niña states. The hatching indicates statistically significant differences between El Nino/La Nina states and Neutral states indicated by a two-sample T-test with a significance threshold of 0.05. }
\label{fig:enso_comp_ocean}
\end{figure}

We now investigate to what extent Ola's  simulated ENSO ocean temperature anomalies exhibit realistic three-dimensional thermal structure. Fig \ref{fig:enso_comp_ocean} shows the averaged SST anomaly and the upper ocean temperature anomaly in the equatorial Pacific from the hundreds of El Niño and La Niña events simulated by the Ola model in month 4 after initialization. For comparison, the corresponding El Niño and La Niña composites were generated from the ERA5 and UFS-replay datasets (ground truth) during the same out-of-sample validations years 2017 to 2021. 

Within the equatorial Pacific, realistic ENSO composites are found. The horizontal extent and the magnitude of the SST anomaly in modeled El Niño and La Nina events (Figure~\ref{fig:enso_comp_ocean} (a), (b)) resemble the observations (Figure~\ref{fig:enso_comp_ocean}(c), (d)). This includes the predominance of the Central Pacific category of El Nino during the past decade. Beneath the ocean surface the simulated (Figure~\ref{fig:enso_comp_ocean} (e), (f)) vertical and zonal locations and extents of the potential temperature anomalies associated with ENSO also skillfully resemble the observations (Figure~\ref{fig:enso_comp_ocean} (g), (h)). For instance during El Nino, whereas ocean surface temperature anomalies maximize in the Central Pacific, sub-surface temperature anomalies at a depth of 50 m maximize in the Eastern Pacific, consistent with the location of the thermocline and implicitly learned shoaling dynamics in the vicinity of its time-mean depth.

Some apparent biases of Ola are also evident in the ENSO composite. Within the Pacific, its El Nino and La Nina SST anomalies could be viewed as too heavily weighted towards the Central Pacific than the observations, consistent with too much subtropical control relative to tropical control. Beyond the tropical Pacific, opposite signed Atlantic and Indian Ocean SST anomalies are also apparent (Figure~\ref{fig:enso_comp_ocean} (b) and (d)). However, it is difficult to fairly assess these long-range teleconnections associated with ENSO, which are inherently noisy compared to their tropical origins, such that the difference in sampling of the many simulated vs. relatively few observed ENSO events becomes an increasingly confounding factor (shown by the insignificant/unhatched region in Figure~\ref{fig:enso_comp_ocean} (d)). That is, there are hundreds of simulated ENSO events yet only a handful observed events in the comparison. 

Nonetheless, shorter-range teleconnections from the Central Pacific to the Western US appear potentially realistic during the El Nino phase - as evidenced by appropriate SST anomalies to the US Southwest corridor and the Pacific Northwest Aleutian Low. This hypothesis deserves further scrutiny. 

\begin{figure}
    \centering
    \includegraphics[scale=0.60]{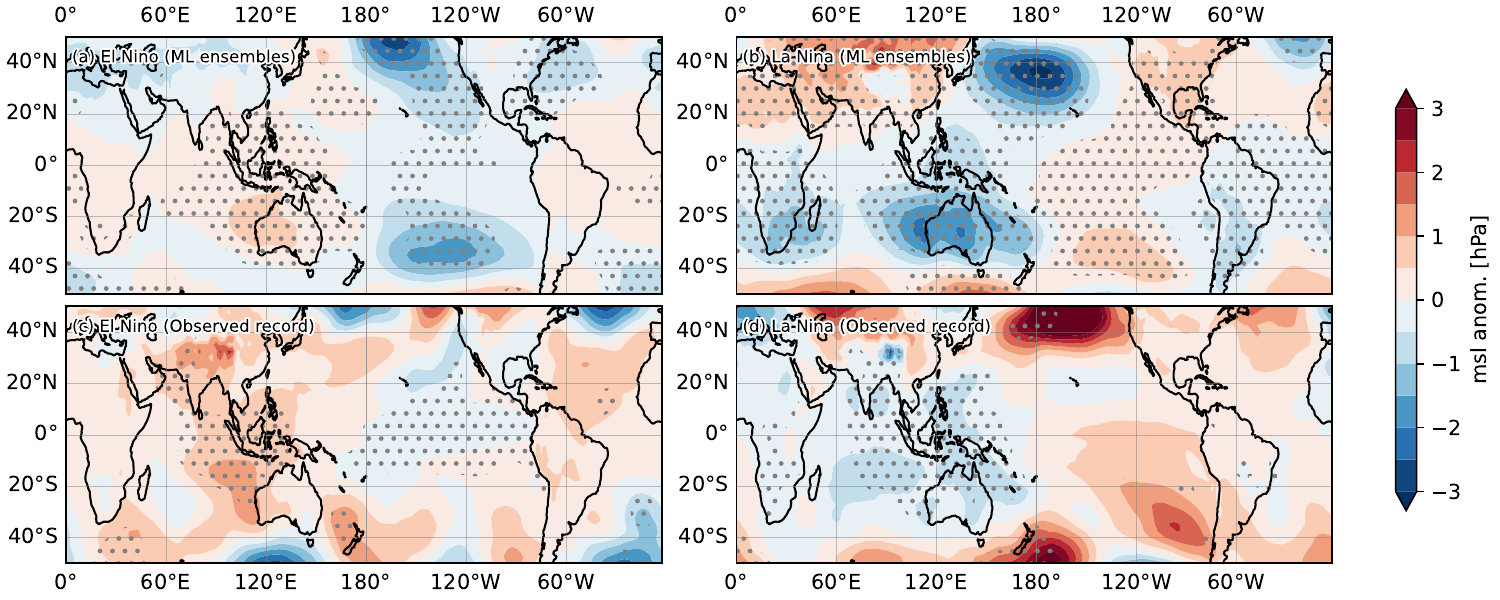}
    \caption{Same as Figure~\ref{fig:enso_comp_ocean} but for  Mean Sea Level Pressure (MSLP) anomaly during El Niño and La Niña events. The top row (a,b) shows the average MSLP anomaly from the Ola coupled ESM at a lead time of 4 months, and the bottom row (c,d) shows the groud truth (ERA5). }
    \label{fig:enso_comp_atmos}
\end{figure}

It is important to investigate whether the atmospheric state is appropriately consistent with the oceanic state during ENSO extrema. Within the atmosphere, we visualize the composite Mean Sea Level Pressure (MSLP) during the El Niño and La Niña phases from simulations of the Ola model (Figure~\ref{fig:enso_comp_atmos}). During the El Niño phase, Ola reproduces the expected zonal MSLP gradient with a high-pressure anomaly in the equatorial Indo-Pacific and a corresponding low-pressure anomaly in the Eastern Pacific, and vice versa during La Nina, consistent with the classical Southern Oscillation. We caution that the atmosphere composites are noisy and thus harder to interpret given the short evaluation period of 5 years. 

Finally, to qualitatively study the transient dynamics in the Ola coupled ESM, we look at two case studies, one showing the onset of the El Niño event starting in May 2018 and another showing the onset of a La Niña event starting in July 2020. Recall from Figure~\ref{fig:enso_plume} that for these initialization dates, Ola predictions tended to generally agree on ENSO phase transition dynamics, within the ensemble spread, such that ensemble averaging is warranted to get a sense of the composite unsteady evolution of each event. Some disagreement is to be expected even in a perfect forecasting system given that one realization of nature is compared to a composite of predicted events, such that our assessment is mostly on the plausibility of the transient evolution in Ola, and whether a reasonably diverse range of ENSO dynamical pathways, including a mixture of tropical vs. subtropical control, can be simulated.

\begin{figure}
    \centering
    \includegraphics[width=0.9\textwidth]{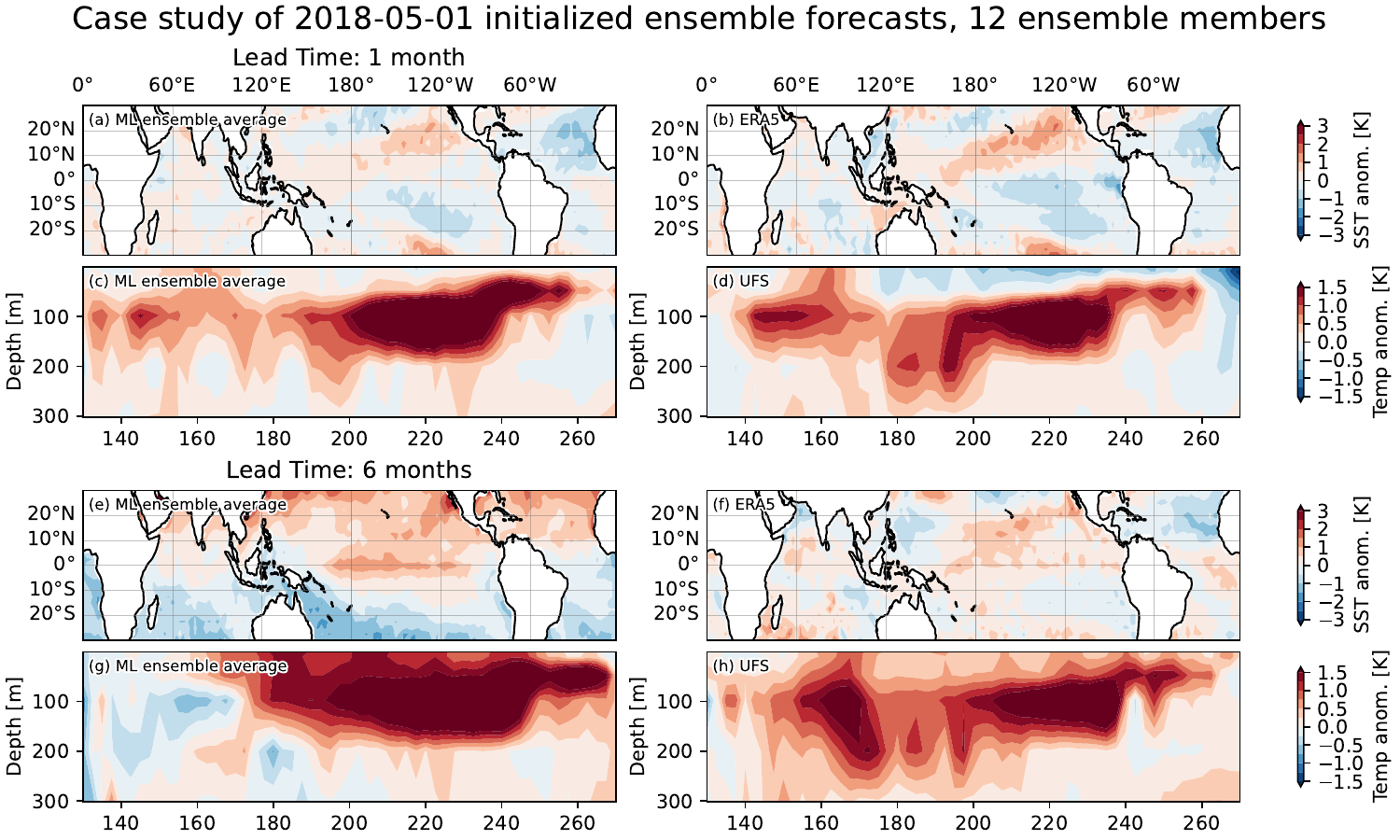}
    \caption{The ensemble mean reforecasts by Ola initialized in 2018/05.  (a) and (c) show the SST  and equatorial Pacific upper ocean temperature anomalies at one month lead time, while  (e) and (g) show the same variables but at 6 months lead time. The results are averaged over 12 ensembles initialized with LEF method (see details in Sec. \ref{sec:LEF}) with biases correction (see details in Sec. \ref{sec:app_bias}). The corresponding observational records ((b), (d), (f) and (h)) are shown in the right column for comparison. }
    \label{fig:case_study_elnino}
\end{figure}

Figure~\ref{fig:case_study_elnino} shows the development of  2018/2019 El Niño and compares the ensemble mean Ola simulation to the observation. Five months ahead of El Niño, a preceding warm anomaly occurs in the subsurface of the equatorial central Pacific region (Figure~\ref{fig:case_study_elnino}(c) vs. (d); 200-230 deg E), consistent with the arrival from the west of a Kelvin wave that often precedes the onset of El Nino (Figure~\ref{fig:eqwave} and Figure~\ref{fig:app_ocn_wave_201804}). This equatorial warm water volume has been understood as an important predictor for El Niño events according to the recharge oscillator theory \cite{jin1997recharge,meinen2000observations} and can be viewed as part of the ``tropical control'' on El Nino dynamics. The tilted thermal anomaly in the depth-longitude plane is a signal that the associated displacement of  the thermocline is captured by the modeling system. Meanwhile, at the surface, another potential control on ENSO is evident via subtropical influence, via the anomalous warm SST that stretches from the southwestern coast of North America to the Central Pacific (Figure~\ref{fig:case_study_elnino}(a,b)). This ``subtropical control'' can sometimes overwhelm the influence of tropical control leading to a Central Pacific type of El Niño rather than the Eastern Pacific type of El Niño \cite{yu2013identifying}. The Central Pacific type of El Niño features more warming in the Central Pacific than the  Eastern Pacific, which is different from the Eastern Pacific type of El Niño. 

Encouragingly, five months later (6 months into the coupled simulations) a mixture of both tropical and subtropical control can be seen to have played a role in Ola (Figure~\ref{fig:case_study_elnino}(e)), as also occurred in the observed event (Figure~\ref{fig:case_study_elnino}(f)). In the Ola composite, the subtropical pathway happened to produce an especially strong influence for this event, consistent with larger subsurface anomalies in the Central Pacific relative to what was observed  (Figure~\ref{fig:case_study_elnino}(g)\&(h)), overwhelming tropical control and leading to a strongly Central Pacific type of El Nino event.  

\begin{figure}
    \centering
    \includegraphics[width=0.9\textwidth]{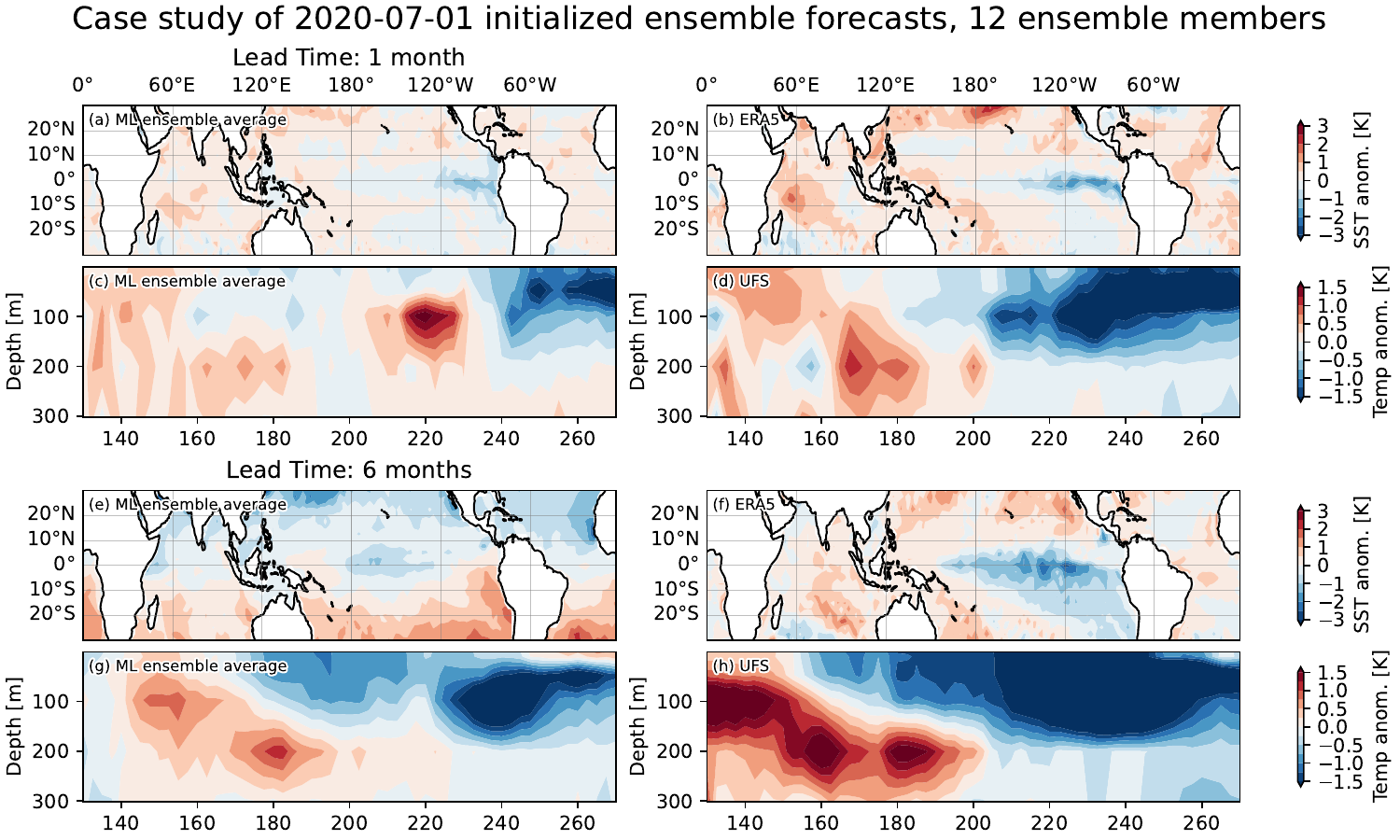}
    \caption{Same as Figure~\ref{fig:case_study_elnino} but for reforecasts initialized in 2020/07.}
    \label{fig:case_study_lanina}
\end{figure}

Ola also predicted the 2020/2021 La Niña with a smaller amplitude for the forecast initiated in July 2020 (Figure~\ref{fig:case_study_lanina}(a)\&(b)). Similar to El Niño, before the onset of La Niña, there is also a clear subsurface anomaly but in the opposite sign (Figure~\ref{fig:case_study_elnino}(c)\&(d)). In this case, a cold anomaly forms in the eastern equatorial region and extends to the central Pacific in both the simulation and observation (Figure~\ref{fig:case_study_elnino}(g)\&(h)). Compared to the observation, Ola simulated La Niña event at the 6-month lead time is weaker than the observation (Figure~\ref{fig:case_study_elnino}(e)\&(f)). The simulated largest signal is more towards the central Pacific than the observation.  We are not certain about the cause of biases in the Ola simulation. One possibility is the lack of ocean currents in this data-driven model, which is critical for the ENSO development \cite{Su2009ensoadvection}.

\section{Discussion, Limitations, and Future Work}\label{sec:discussion}
We present the Ola model, a pioneering AI/ML coupled Earth System Model. Our model can generate a 6-month forecast of the coupled atmosphere-ocean system in less than 1 minute with one A100 GPU. This offers a significant speed-up of several orders of magnitude compared with process-based numerical weather and climate models.  

By coupling the ML atmosphere model and the ML ocean model, the Ola model is able to produce convincing simulations and forecasts of ENSO, the dominant interannual variability in Earth's climate and the leading source of the predictability in seasonal to interannual time scale. Analysis of Ola suggests learned oceanic and coupled physics on seasonal timescales, as evidenced by correct subsurface potential temperature vertical and horizontal structures associated with ENSO anomalies, oceanic equatorial waves that propagate at appropriate phases across a range of tropical latitude bands, and overall ENSO variability within the Pacific Basin. We encourage formal tests of the hypothesis that models like Ola can learn coupled and ocean dynamics, as has been done for atmosphere-only ML models \cite{hakim2023dynamical}.

Fortuitously, within the limited evaluation performed, the Ola model exhibits substantially smaller systematic biases in the tropics compared to a conventional process-based coupled model, GFDL-SPEAR. Indeed, the lack of tropical drift in Ola is remarkable and one of the enticing reasons to explore ML simulation, given that physically based coupled modeling systems have been prone to stubborn biases such as the double ITCZ and equatorial cold tongue for decades \cite{tian2020double,chen2017ensobias, capotondi2015climate, li2014modelbias} 

An important limitation of our current model is the occurrence of substantial drifts at higher latitudes (see Appendix \ref{sec:app_bias}), despite the absence of tropical drift. These drifts cause dynamical instabilities for rollout periods longer than six months.. We do not view this as an inherent limitation of coupled ML models in general or the SFNO in particular. Previous studies, such as ~\cite{watt2023ace}, have demonstrated that stable simulations as long as a century are feasible with ML models, given adequate data and sufficient tuning. Understanding the origins of such long-term drift attractors in autoregressive ML, and how to control them, is an open area deserving of systematic empirical testing. 

While such a comprehensive survey is beyond the scope of this proof-of-concept study, we acknowledge that some of our empirical ML choices, such as the timestep of our atmospheric model, did have an influence on controlling the severity of our long-term coupled bias attractor. Early approaches in this research attempted to couple an SST-only based ocean to a default SFNO atmosphere analogous to that in \cite{bonev2023spherical}, which used a 6-hour timestep, but this resulted in a model that unsatisfyingly converged to a permanent El Niño state within 6 model months. Whether this is a meaningful signal or an effect of empirical ML noise is unknown. Contrary to our experience, Creswell-Clay and colleagues (personal communications) have observed initial stabilizing effects from similar ocean coupling attempts with their SST-based ocean ML model and a DLWP-HPX atmosphere\cite{weyn2021sub} even when using a 6-hour atmospheric model time step. This suggests significant empirical variability, and the need for systematic model intercomparisons to fully understand and optimize long-term mean state drift, an important topic towards advancing coupled ML towards climate simulation.

Meanwhile, it is clear that coupled atmosphere-ocean ML/AI simulation is already possible on seasonal timescales, with exciting potential and obvious extensions worthy of further work. It would be worth training future iterations of this model with increasingly ambitious ocean state vectors that better resolve oceanic physics. The UFS-replay and publicly available ocean reanalysis datasets have a short period of record beginning in 1994 due to historical limitations of observing systems. This presents an impediment to training ML models, which typically improve when trained on more data. We encourage the use of future datasets with climate model simulations for pre-training ML models and using historical re-analysis data for finetuning. For extensions to future climate prediction, it is logical to expect that vast training libraries will eventually become necessary, comprising many thousands of years of ocean simulation output to sample the internal variability and forced dynamics of slower than seasonal components of the ocean circulation, such as subtropical gyres and the Atlantic meridional overturning circulation. We are encouraged that hybrid AI/physics climate modeling thought leaders are actively lobbying national simulation centers to develop and publish such datasets in formats easily accessible to the community (Laure Zanna and Chris Bretherton; personal communication).

Meanwhile, we present this model as a part of a research effort towards extending the capabilities of AI/ML models beyond medium-range, atmosphere-only weather forecasting. We acknowledge that more rigorous evaluation and further development are necessary before the model can provide useful operational seasonal forecasts. We hope that our work provides some benchmarks and helps inform the development of such AI/ML models at operational NWP centers. Towards this end, we intend to make our training and inference code publicly available along with trained model weights and data.

\section{Data and Methods}\label{sec:methods}
\subsection{Data-driven Atmosphere and Ocean Model with SFNO}\label{sec:model_config}
We separately develop the ocean and atmosphere components of the Ola coupled ESM based on the Spherical Fourier Neural Operator architecture. This architecture has previously been used to develop an atmospheric model exhibiting competitive medium-range forecast skill relative to SOTA ML and NWP models~\cite{bonev2023spherical}. Moreover, the architecture exhibits long-term stable dynamics for 10 to 100 years when trained on a sufficient quantity of data~\cite{watt2023ace}. We use the publicly available NVIDIA modulus-makani~\footnote{https://github.com/NVIDIA/modulus-makani} package to train our atmosphere and ocean models with appropriate modifications for accommodating an ocean state vector.

Let $A_t$ and $O_t$ represent the instantaneous states of the atmosphere and the ocean at time $t$ respectively. The states are composed of stacked 2-dimensional atmospheric and oceanic variables which sample the 3-dimensional and multivariate structure of the atmosphere and the ocean. We define ML atmosphere model ($F^{A}$) and ocean model ($F^{O}$) with the SFNO architecture such that

\begin{align}\label{eq:coupling}
    A_{t+\Delta t} \approx F^{A}\left[ A_t, \tilde{O}_t, Z_t \right], \\
    O_{t+\Delta t^\prime} \approx F^{O}\left[O_t,  \tilde{A}_t, Z_t \right] 
\end{align}

where $\tilde{A}_t$ ($\subseteq A_t$) and $\tilde{O}_t$ ($\subseteq O_t$),  which are a subset of variables contained in atmospheric states ($A_t$) and oceanic states $O_t$, are used as the boundary conditions for the ocean and atmosphere model (Figure~\ref{fig:coupling-schematic}). The variables for the state vectors and boundary conditions are listed in the Appendix \ref{sec:app_vars}. Both models also use an internally computed estimate of the cosine of the solar zenith angle at each grid point at time $t$ denoted $Z_t$. The solar zenith angle provides the models with contextual information about the annual cycle. We first train the $F^A$ and $F^O$ separately using assimilated observation datasets (see Sec. \ref{sec:training_data}) and a mean-squared loss function defined appropriately on the sphere. Then, the atmosphere and ocean models ($F^A$ and $F^O$) are coupled to perform the forecast (see Sec. \ref{sec:coupling}).  

\begin{figure}
    \centering
    \includegraphics[width=0.9\textwidth, trim=0 2.5cm 0 2.5cm, clip]{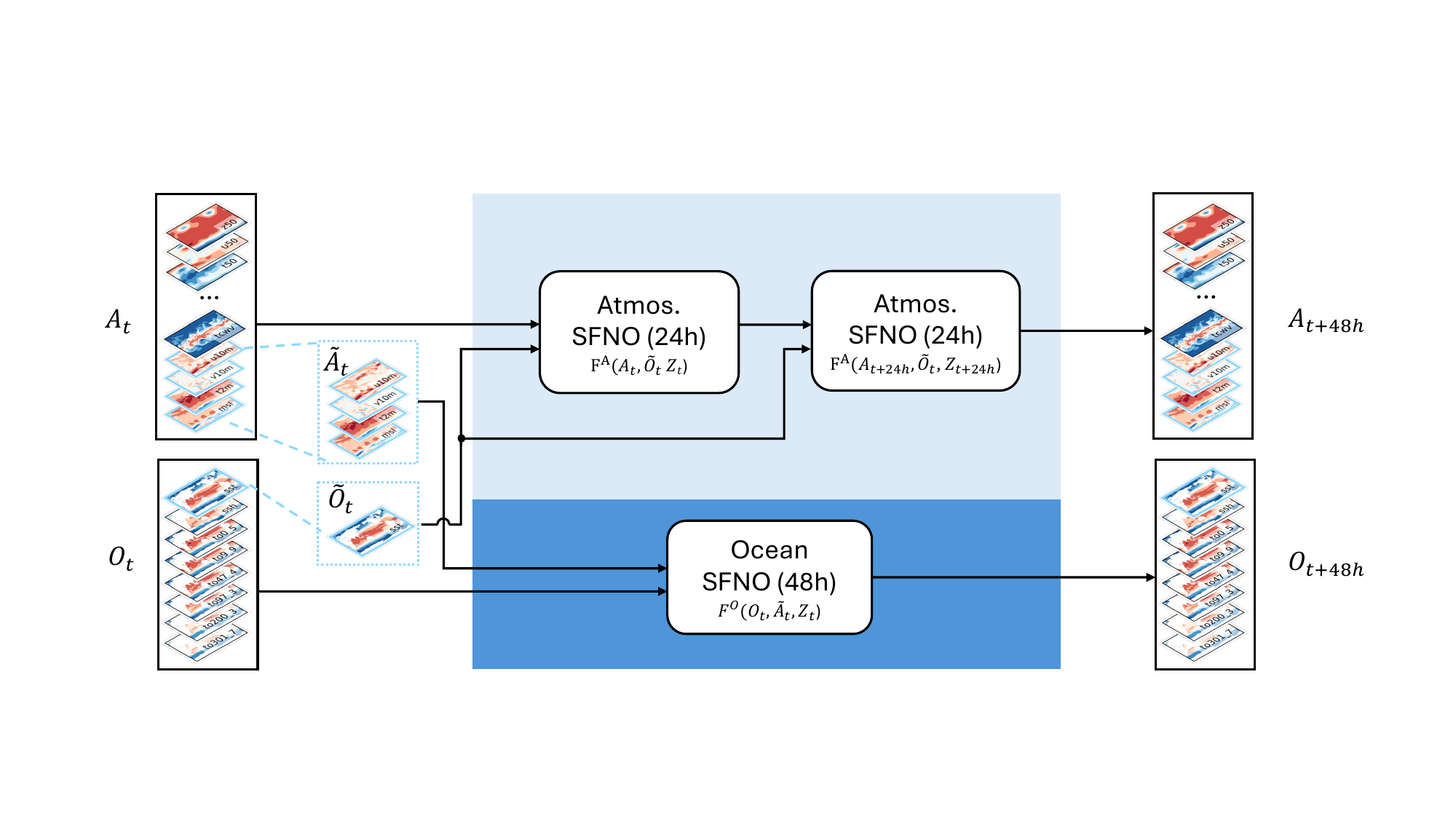}
    \caption{Our coupled ESM consists of separately trained atmosphere and ocean models. The atmosphere and ocean models are based on an SFNO architecture. The atmosphere model uses a timestep of 24 hours and the ocean model uses a timestep of 48 hours.}
    \label{fig:coupling-schematic}
\end{figure}

\subsection{Training Data} \label{sec:training_data}

For the atmosphere, we keep convention with recent literature by training on the ERA5 reanalysis dataset~\cite{hersbach2020era5}, a gridded estimate of the observed atmospheric state, at its native ($ 0.25^{\circ} \text{ latitude} \times 0.25^{\circ} \text{ longitude} $) resolution. Of the 137 pressure levels available at hourly frequency, we train the atmosphere component of the model ($F^A$) using a subset of variables at 13 different pressure levels spanning the surface to the top of the troposphere (see Appendix~\ref{sec:app_vars} for a list of the atmospheric state variables). The ERA5 fields are temporally sampled using an interval of 6 hours at 0000, 0600 1200 and 1800 UTC for all days in the training period which is chosen to be from 1980 to 2016. The timestep for training is 24 hours for the atmosphere model. A few samples from 2017 were used for validation during training wherein the 1-step validation loss was calculated to visualize convergence of training. Since our evaluation protocol looks at very long rollouts of 100s of steps, we do not exclude 2017 from the evaluation set. We use the data from 2017 onwards for testing.

The ocean component of the model $F^O$ is trained on the UFS-replay dataset~\cite{ufsreplay} developed by the National Oceanic and Atmosphere Administration (NOAA) using their next-generation Unified Forecast System (UFS) model and a nudging procedure that synchronizes the oceanic model to the ERA5 historical record. The UFS-replay dataset offers a good high-resolution, densely sampled historical record of the ocean system. The reanalysis-like UFS-replay dataset incorporates a physical prior from the modern UFS model with a publicly accessible analysis-ready cloud-optimized store. We utilize only a fraction of the prognostic ocean variables contained in the UFS-replay dataset to train our model. Specifically, we choose to include the Sea Surface Temperature (SST), the Sea Surface Height (SSH), and the potential temperature at eight vertical levels from 0.5m below the surface to 300m below the surface (see Appendix~\ref{sec:app_vars} for a list of all the ocean state variables used for training). The UFS-replay fields are temporally sampled with an interval of 6 hours at 0000, 0600, 1200, 1800 UTC for all days in the training period which is chosen to be from 1994 to 2016. The data from 2017 onwards is reserved for testing.

Since the ocean variables are undefined over land, we utilize a binary mask appropriate to each ocean level at the input and within the training loss function. For any ocean vector in the input, we apply the binary mask which sets all undefined values to zero before passing the vector to the model. For all predicted ocean fields that appear in the output of a model, we apply the binary mask to the model output and the target before computing the mean-squared loss function.

\subsection{Coupling procedure} \label{sec:coupling}

We model the interaction between the ocean and atmosphere as being mediated by a handful of surface and near-surface variables. In Eqs.~\ref{eq:coupling} the atmosphere model is forced by the subset $\tilde{O}$ of the ocean state vector consisting of only the sea surface temperature (SST). This recognizes that the main role of ocean-to-atmosphere coupling in tropical atmospheric dynamics is through the local modulation of surface fluxes, like latent heat fluxes, and of flow from horizontal pressure gradients, which are primarily tied to SST in the ocean component.

The ocean model is forced by a subset $\tilde{A}$ of the atmospheric variables consisting of the near-surface zonal and meridional wind velocities ( $10u$, $10v$ ), the 2 meter temperature ($2t$), and the mean sea-level pressure ($mslp$). This recognizes that the ocean is mechanically forced by atmospheric surface winds, and that the surface wind and surface temperature contrast are the first-order controls on latent and sensible heat fluxes, which are modeled implicitly as in both the atmospheric and oceanic components.

We also choose different timesteps for the atmosphere and ocean components. The atmosphere model is trained with a timestep of 24 hours while the ocean model is trained with a timestep of 48 hours. Figure ~\ref{fig:coupling-schematic} illustrates our coupling strategy for running the models in autoregressive coupled inference mode. At $t=0$, the models are initialized using reanalysis data from UFS-replay and ERA5. The ocean and atmosphere states are then updated autoregressively following Eqs~(1, 2) with the ocean state being updated at a timestep of 48 hours and the atmosphere state updated using a timestep of 24 hours. We use the tooling provided by the earth2mip project\footnote{https://github.com/NVIDIA/earth2mip} for orchestrating the coupled inference and conveniently saving the required outputs for analysis.

\subsection{Forecast with lagged ensembles}\label{sec:LEF}

Ensemble forecasts are crucial for assessing the variability in weather forecasts due to the chaotic physics of the atmosphere-ocean system. Physics-based NWP models generate ensemble forecasts by perturbing the initial conditions and model parameters to obtain a Monte Carlo sample of forecasts from a given initial condition. Finding optimal perturbations to correctly sample the uncertainty in initial conditions and also promote dispersion in modeled dynamics towards well-calibrated probabilistic predictions is a very involved problem that requires much tuning and experimentation, beyond the scope of this proof of concept.

We adopt the Lagged Ensemble Forecasting (LEF) method \cite{Hoffman1983-sy} to generate long-range ensemble forecasts with the Ola model. LEF forms an ensemble based on a series of unperturbed deterministic forecasts spawned at varying initial times but assessed at a common validation time. In this work, we initialize the model every month allowing the model to evolve freely for 6 months following initialization. The model is initialized every 6 hours in the first 3 days of each month to create 12 ensemble members for every forecast. While LEF was superseded with other modern, parametrized methods, it has many advantages. it allows efficient sampling of the probabilistic character of prototype forecast models in a parameter-free manner with minimal experimentation and avoiding confounding factors that allow a clean comparison to an operational baseline~\cite{brenowitz2024practical}. LEF has been used for ensemble generation in recent work on S2S and seasonal forecasting research~\cite{vitart2021lagged,  trenary2018monthly}.

\section*{Acknowledgments}
We thank Nathaniel Cresswell-Clay for sharing his coupled ML Atmosphere-Ocean  model. We thank Prof. Laure Zanna for helpful comments. We thank Prof. Jin-Yi Yu for helpful guidance on ENSO evaluation. We thank the NVIDIA Modulus team for software support. We thank the ECMWF and NCEP/NOAA for open sharing of data and fundamental research which enabled our work. We acknowledge the agencies that support the NMME-Phase II system, and we thank the climate modeling groups (Environment Canada, NASA, NCAR, NOAA/GFDL, NOAA/NCEP, and University of Miami) for producing and making available their model output. NOAA/NCEP, NOAA/CTB, and NOAA/CPO jointly provided coordinating support and led development of the NMME-Phase II system.

\bibliographystyle{unsrt}
\bibliography{references}

\begin{thebibliography}{10}

\bibitem{bonev2023spherical}
Boris Bonev, Thorsten Kurth, Christian Hundt, Jaideep Pathak, Maximilian Baust, Karthik Kashinath, and Anima Anandkumar.
\newblock Spherical fourier neural operators: Learning stable dynamics on the sphere.
\newblock In {\em International conference on machine learning}, pages 2806--2823. PMLR, 2023.

\bibitem{bi2023accurate}
Kaifeng Bi, Lingxi Xie, Hengheng Zhang, Xin Chen, Xiaotao Gu, and Qi~Tian.
\newblock Accurate medium-range global weather forecasting with 3d neural networks.
\newblock {\em Nature}, 619(7970):533--538, 2023.

\bibitem{lam2022graphcast}
Remi Lam, Alvaro Sanchez-Gonzalez, Matthew Willson, Peter Wirnsberger, Meire Fortunato, Ferran Alet, Suman Ravuri, Timo Ewalds, Zach Eaton-Rosen, Weihua Hu, et~al.
\newblock Graphcast: Learning skillful medium-range global weather forecasting.
\newblock {\em arXiv preprint arXiv:2212.12794}, 2022.

\bibitem{pathak2022fourcastnet}
Jaideep Pathak, Shashank Subramanian, Peter Harrington, Sanjeev Raja, Ashesh Chattopadhyay, Morteza Mardani, Thorsten Kurth, David Hall, Zongyi Li, Kamyar Azizzadenesheli, et~al.
\newblock Fourcastnet: A global data-driven high-resolution weather model using adaptive fourier neural operators.
\newblock {\em arXiv preprint arXiv:2202.11214}, 2022.

\bibitem{karlbauer2023advancing}
Matthias Karlbauer, Nathaniel Cresswell-Clay, Dale~R Durran, Raul~A Moreno, Thorsten Kurth, and Martin~V Butz.
\newblock Advancing parsimonious deep learning weather prediction using the healpix mes.
\newblock {\em Authorea Preprints}, 2023.

\bibitem{chen2023fengwu}
Kang Chen, Tao Han, Junchao Gong, Lei Bai, Fenghua Ling, Jing-Jia Luo, Xi~Chen, Leiming Ma, Tianning Zhang, Rui Su, et~al.
\newblock Fengwu: Pushing the skillful global medium-range weather forecast beyond 10 days lead.
\newblock {\em arXiv preprint arXiv:2304.02948}, 2023.

\bibitem{keisler2022forecasting}
Ryan Keisler.
\newblock Forecasting global weather with graph neural networks.
\newblock {\em arXiv preprint arXiv:2202.07575}, 2022.

\bibitem{rasp2023weatherbench}
Stephan Rasp, Stephan Hoyer, Alexander Merose, Ian Langmore, Peter Battaglia, Tyler Russel, Alvaro Sanchez-Gonzalez, Vivian Yang, Rob Carver, Shreya Agrawal, et~al.
\newblock Weatherbench 2: A benchmark for the next generation of data-driven global weather models.
\newblock {\em arXiv preprint arXiv:2308.15560}, 2023.

\bibitem{brenowitz2024practical}
Noah~D Brenowitz, Yair Cohen, Jaideep Pathak, Ankur Mahesh, Boris Bonev, Thorsten Kurth, Dale~R Durran, Peter Harrington, and Michael~S Pritchard.
\newblock A practical probabilistic benchmark for ai weather models.
\newblock {\em arXiv preprint arXiv:2401.15305}, 2024.

\bibitem{bouallegue2024aifs}
Zied~Ben Bouallegue, Mihai Alexe, Matthew Chantry, Mariana Clare, Jesper Dramsch, Simon Lang, Christian Lessig, Linus Magnusson, Ana~Prieto Nemesio, Florian Pinault, et~al.
\newblock Aifs--ecmwf’s data-driven probabilistic forecasting.
\newblock Technical report, Copernicus Meetings, 2024.

\bibitem{brankovic1994predictability}
{\v{C}}edomir Brankovi{\'c}, TN~Palmer, and L~Ferranti.
\newblock Predictability of seasonal atmospheric variations.
\newblock {\em Journal of Climate}, 7(2):217--237, 1994.

\bibitem{walker1924correlations1}
GT~WALKER.
\newblock Correlation in seasonal variations of weather. ix: A preliminary study of world-weather.
\newblock {\em Mem. Indian Meteor. Dep.}, 24:275--332, 1924.

\bibitem{mcphaden2006enso}
Michael~J McPhaden, Stephen~E Zebiak, and Michael~H Glantz.
\newblock Enso as an integrating concept in earth science.
\newblock {\em science}, 314(5806):1740--1745, 2006.

\bibitem{bjerknes1969atmospheric}
Jakob Bjerknes.
\newblock Atmospheric teleconnections from the equatorial pacific.
\newblock {\em Monthly weather review}, 97(3):163--172, 1969.

\bibitem{cane2005enso}
Mark~A Cane.
\newblock The evolution of el ni{\~n}o, past and future.
\newblock {\em Earth and Planetary Science Letters}, 230(3-4):227--240, 2005.

\bibitem{cane1985theory}
Mark~A Cane and Stephen~E Zebiak.
\newblock A theory for el ni{\~n}o and the southern oscillation.
\newblock {\em Science}, 228(4703):1085--1087, 1985.

\bibitem{cane1986experimental}
Mark~A Cane, Stephen~E Zebiak, and Sean~C Dolan.
\newblock Experimental forecasts of el nino.
\newblock {\em Nature}, 321(6073):827--832, 1986.

\bibitem{weyn2021sub}
Jonathan~A Weyn, Dale~R Durran, Rich Caruana, and Nathaniel Cresswell-Clay.
\newblock Sub-seasonal forecasting with a large ensemble of deep-learning weather prediction models.
\newblock {\em Journal of Advances in Modeling Earth Systems}, 13(7):e2021MS002502, 2021.

\bibitem{arcomano2022hybrid}
Troy Arcomano, Istvan Szunyogh, Alexander Wikner, Jaideep Pathak, Brian~R Hunt, and Edward Ott.
\newblock A hybrid approach to atmospheric modeling that combines machine learning with a physics-based numerical model.
\newblock {\em Journal of Advances in Modeling Earth Systems}, 14(3):e2021MS002712, 2022.

\bibitem{watt2023ace}
Oliver Watt-Meyer, Gideon Dresdner, Jeremy McGibbon, Spencer~K Clark, Brian Henn, James Duncan, Noah~D Brenowitz, Karthik Kashinath, Michael~S Pritchard, Boris Bonev, et~al.
\newblock Ace: A fast, skillful learned global atmospheric model for climate prediction.
\newblock {\em arXiv preprint arXiv:2310.02074}, 2023.

\bibitem{kochkov2023neural}
Dmitrii Kochkov, Janni Yuval, Ian Langmore, Peter Norgaard, Jamie Smith, Griffin Mooers, James Lottes, Stephan Rasp, Peter D{\"u}ben, Milan Kl{\"o}wer, et~al.
\newblock Neural general circulation models.
\newblock {\em arXiv preprint arXiv:2311.07222}, 2023.

\bibitem{xiong2023ai}
Wei Xiong, Yanfei Xiang, Hao Wu, Shuyi Zhou, Yuze Sun, Muyuan Ma, and Xiaomeng Huang.
\newblock Ai-goms: Large ai-driven global ocean modeling system.
\newblock {\em arXiv preprint arXiv:2308.03152}, 2023.

\bibitem{subel2024building}
Adam Subel and Laure Zanna.
\newblock Building ocean climate emulators.
\newblock {\em arXiv preprint arXiv:2402.04342}, 2024.

\bibitem{chattopadhyay2023oceannet}
Ashesh Chattopadhyay, Michael Gray, Tianning Wu, Anna~B Lowe, and Ruoying He.
\newblock Oceannet: A principled neural operator-based digital twin for regional oceans.
\newblock {\em arXiv preprint arXiv:2310.00813}, 2023.

\bibitem{saha2014ncep}
Suranjana Saha, Shrinivas Moorthi, Xingren Wu, Jiande Wang, Sudhir Nadiga, Patrick Tripp, David Behringer, Yu-Tai Hou, Hui-ya Chuang, Mark Iredell, et~al.
\newblock The ncep climate forecast system version 2.
\newblock {\em Journal of climate}, 27(6):2185--2208, 2014.

\bibitem{johnson2019seas5}
Stephanie~J Johnson, Timothy~N Stockdale, Laura Ferranti, Magdalena~A Balmaseda, Franco Molteni, Linus Magnusson, Steffen Tietsche, Damien Decremer, Antje Weisheimer, Gianpaolo Balsamo, et~al.
\newblock Seas5: the new ecmwf seasonal forecast system.
\newblock {\em Geoscientific Model Development}, 12(3):1087--1117, 2019.

\bibitem{lu2020gfdl}
Feiyu Lu, Matthew~J Harrison, Anthony Rosati, Thomas~L Delworth, Xiaosong Yang, William~F Cooke, Liwei Jia, Colleen McHugh, Nathaniel~C Johnson, Mitchell Bushuk, et~al.
\newblock Gfdl's spear seasonal prediction system: Initialization and ocean tendency adjustment (ota) for coupled model predictions.
\newblock {\em Journal of Advances in Modeling Earth Systems}, 12(12):e2020MS002149, 2020.

\bibitem{chen2017ensobias}
Chen Chen, Mark~A Cane, Andrew~T Wittenberg, and Dake Chen.
\newblock Enso in the cmip5 simulations: Life cycles, diversity, and responses to climate change.
\newblock {\em Journal of Climate}, 30(2):775--801, 2017.

\bibitem{li2014modelbias}
Gen Li and Shang-Ping Xie.
\newblock Tropical biases in cmip5 multimodel ensemble: The excessive equatorial pacific cold tongue and double itcz problems.
\newblock {\em Journal of Climate}, 27(4):1765--1780, 2014.

\bibitem{yu1999links}
Jin-Yi Yu and Carlos~R Mechoso.
\newblock Links between annual variations of peruvian stratocumulus clouds and of sst in the eastern equatorial pacific.
\newblock {\em Journal of Climate}, 12(11):3305--3318, 1999.

\bibitem{song2009convection}
Xiaoliang Song and Guang~Jun Zhang.
\newblock Convection parameterization, tropical pacific double itcz, and upper-ocean biases in the ncar ccsm3. part i: Climatology and atmospheric feedback.
\newblock {\em Journal of Climate}, 22(16):4299--4315, 2009.

\bibitem{Wu2022ColdTongueBias}
Xian Wu, Yuko~M. Okumura, Pedro~N. DiNezio, Stephen~G. Yeager, and Clara Deser.
\newblock The equatorial pacific cold tongue bias in cesm1 and its influence on enso forecasts.
\newblock {\em Journal of Climate}, 35(11):3261 -- 3277, 2022.

\bibitem{wang2020inter}
Chenggong Wang, Yongyun Hu, Xinyu Wen, Chen Zhou, and Jiping Liu.
\newblock Inter-model spread of the climatological annual mean hadley circulation and its relationship with the double itcz bias in cmip5.
\newblock {\em Climate Dynamics}, 55:2823--2834, 2020.

\bibitem{tian2020double}
Baijun Tian and Xinyu Dong.
\newblock The double-itcz bias in cmip3, cmip5, and cmip6 models based on annual mean precipitation.
\newblock {\em Geophysical Research Letters}, 47(8):e2020GL087232, 2020.

\bibitem{song2018roles}
Xiaoliang Song and Guang~J Zhang.
\newblock The roles of convection parameterization in the formation of double itcz syndrome in the ncar cesm: I. atmospheric processes.
\newblock {\em Journal of Advances in Modeling Earth Systems}, 10(3):842--866, 2018.

\bibitem{woelfle2019evolution}
MD~Woelfle, CS~Bretherton, C~Hannay, and R~Neale.
\newblock Evolution of the double-itcz bias through cesm2 development.
\newblock {\em Journal of Advances in Modeling Earth Systems}, 11(7):1873--1893, 2019.

\bibitem{Sprintall2020}
Janet Sprintall, Sophie Cravatte, Boris Dewitte, Yan Du, and Alexander~Sen Gupta.
\newblock {\em ENSO Oceanic Teleconnections}, chapter~15, pages 337--359.
\newblock American Geophysical Union (AGU), 2020.

\bibitem{wang2001unified}
Chunzai Wang.
\newblock A unified oscillator model for the el ni{\~n}o--southern oscillation.
\newblock {\em Journal of Climate}, 14(1):98--115, 2001.

\bibitem{Vallis06}
G.~K. Vallis.
\newblock {\em Atmospheric and Oceanic Fluid Dynamics}.
\newblock Cambridge University Press, Cambridge, U.K., 2006.

\bibitem{jin1997recharge}
Fei-Fei Jin.
\newblock An equatorial ocean recharge paradigm for enso. part ii: A stripped-down coupled model.
\newblock {\em Journal of the Atmospheric Sciences}, 54(7):830--847, 1997.

\bibitem{meinen2000observations}
Christopher~S Meinen and Michael~J McPhaden.
\newblock Observations of warm water volume changes in the equatorial pacific and their relationship to el ni{\~n}o and la ni{\~n}a.
\newblock {\em Journal of Climate}, 13(20):3551--3559, 2000.

\bibitem{yu2013identifying}
Jin-Yi Yu and Seon~Tae Kim.
\newblock Identifying the types of major el ni{\~n}o events since 1870.
\newblock {\em International journal of climatology}, 33(8):2105--2112, 2013.

\bibitem{Su2009ensoadvection}
Jingzhi Su, Renhe Zhang, Tim Li, Xinyao Rong, J-S. Kug, and Chi-Cherng Hong.
\newblock {Causes of the El Ni{\~{n}}o and La Ni{\~{n}}a Amplitude Asymmetry in the Equatorial Eastern Pacific}.
\newblock {\em Journal of Climate}, 23(3):605--617, 2010.

\bibitem{hakim2023dynamical}
Gregory~J Hakim and Sanjit Masanam.
\newblock Dynamical tests of a deep-learning weather prediction model.
\newblock {\em arXiv preprint arXiv:2309.10867}, 2023.

\bibitem{capotondi2015climate}
Antonietta Capotondi, Y~Ham, Andrew Wittenberg, and J~Kug.
\newblock Climate model biases and el ni{\~n}o southern oscillation (enso) simulation.
\newblock {\em US CLIVAR Variations}, 2015.

\bibitem{hersbach2020era5}
Hans Hersbach, Bill Bell, Paul Berrisford, Shoji Hirahara, Andr{\'a}s Hor{\'a}nyi, Joaqu{\'\i}n Mu{\~n}oz-Sabater, Julien Nicolas, Carole Peubey, Raluca Radu, Dinand Schepers, et~al.
\newblock The era5 global reanalysis.
\newblock {\em Quarterly Journal of the Royal Meteorological Society}, 146(730):1999--2049, 2020.

\bibitem{ufsreplay}
Noaa unified forecast system replay reanalysis.
\newblock \url{https://noaa-ufs-gefsv13replay-pds.s3.amazonaws.com/index.html}.
\newblock Accessed: March 2024.

\bibitem{Hoffman1983-sy}
Ross~N Hoffman and Eugenia Kalnay.
\newblock Lagged average forecasting, an alternative to monte carlo forecasting.
\newblock {\em Tellus A}, 35A(2):100--118, March 1983.

\bibitem{vitart2021lagged}
Fr{\'e}d{\'e}ric Vitart and Yuhei Takaya.
\newblock Lagged ensembles in sub-seasonal predictions.
\newblock {\em Quarterly Journal of the Royal Meteorological Society}, 147(739):3227--3242, 2021.

\bibitem{trenary2018monthly}
L~Trenary, T~DelSole, MK~Tippett, and K~Pegion.
\newblock Monthly enso forecast skill and lagged ensemble size.
\newblock {\em Journal of Advances in Modeling Earth Systems}, 10(4):1074--1086, 2018.

\end{thebibliography}
\appendix
\newcounter{AppendixCounter}
\renewcommand{\theAppendixCounter}{\Alph{AppendixCounter}} 
\renewcommand\thefigure{\Alph{AppendixCounter}\arabic{figure}}   

\refstepcounter{AppendixCounter} 
\section*{Appendix \theAppendixCounter: Atmospheric and Oceanic Fields}\label{sec:app_vars}
\setcounter{figure}{0}
List of prognostic variables in the atmosphere and ocean models.

\begin{enumerate}
    \item Atmosphere
    \begin{itemize}
        \item Surface/single level: 10m zonal wind, 10m meridional wind, 2m temperature, surface pressure, total column of integrated water vapor, mean sea level pressure.
        \item Pressure level: Zonal wind (u), Meridional wind (v) Geopotential (z), Temperature (t), Specific Humidity (q) at the following pressure levels in hPa: [50, 100, 150, 200, 250, 300, 400, 500, 600, 700, 850, 925, 1000]
    \end{itemize}
    \item Ocean
    \begin{itemize}
        \item Surface: Sea Surface Temperature (SST), Sea Surface Height (SSH)
        \item Ocean Depth: Potential temperature at the following depths in meters: 
        [0.5, 9.8, 47.3, 97.2, 200.3, 301.7~]
    \end{itemize}
\end{enumerate}

% \clearpage
\refstepcounter{AppendixCounter} 
\section*{Appendix \theAppendixCounter: Ocean wave dynamics}\label{sec:app_waves}
\setcounter{figure}{0}

Simulations showing equatorial ocean wave dynamics in the model evaluation period from 2017 to 2021.

\begin{figure}[h]
    \centering
    \includegraphics[width=0.7\textwidth]{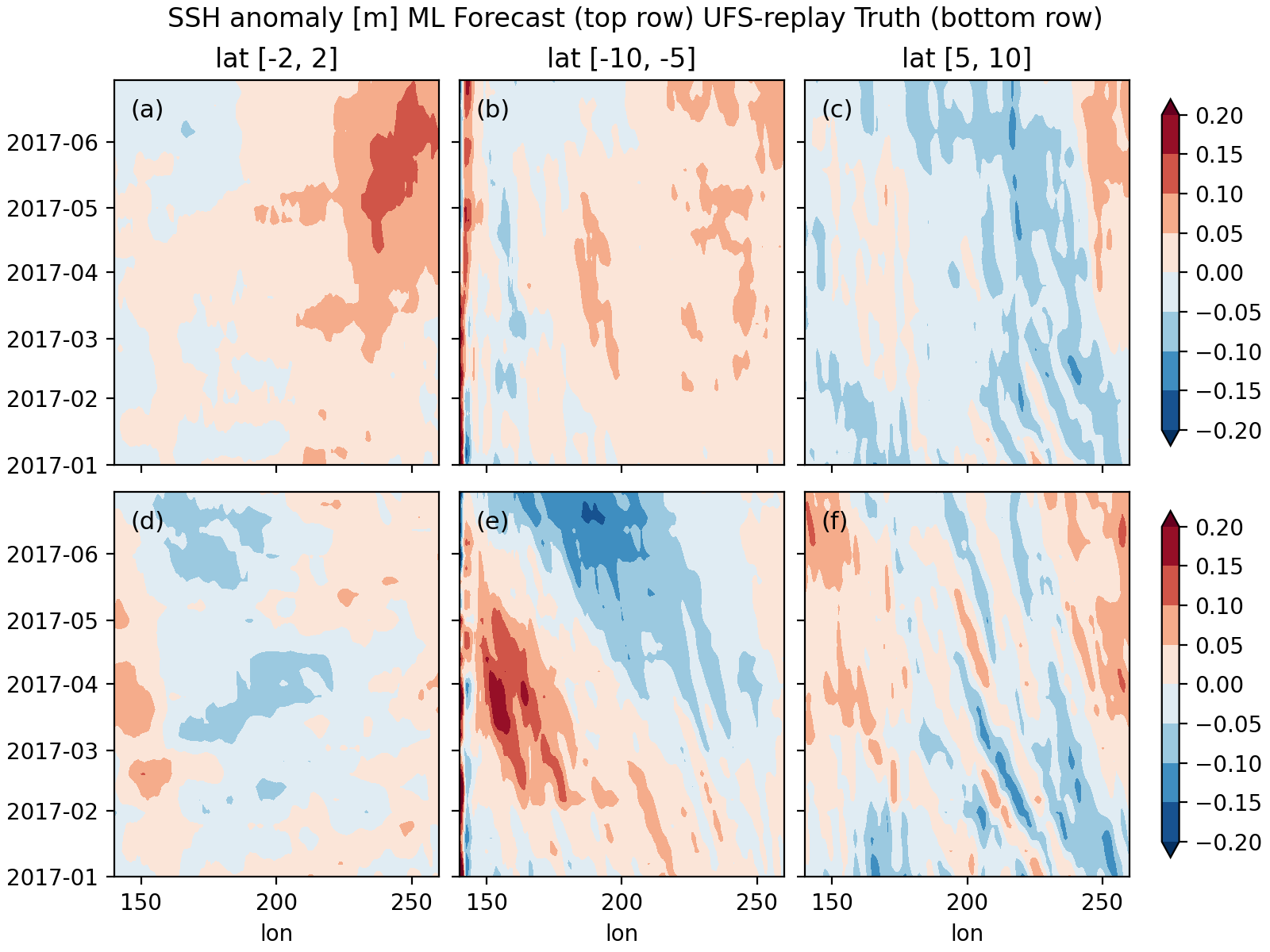}
    \caption{Same as Figure~\ref{fig:eqwave} but for forecast initialized at 2018/01/01-00UTC}
    \label{fig:app_ocn_wave_201701}
\end{figure}

\begin{figure}
    \centering
    \includegraphics[width=0.7\textwidth]{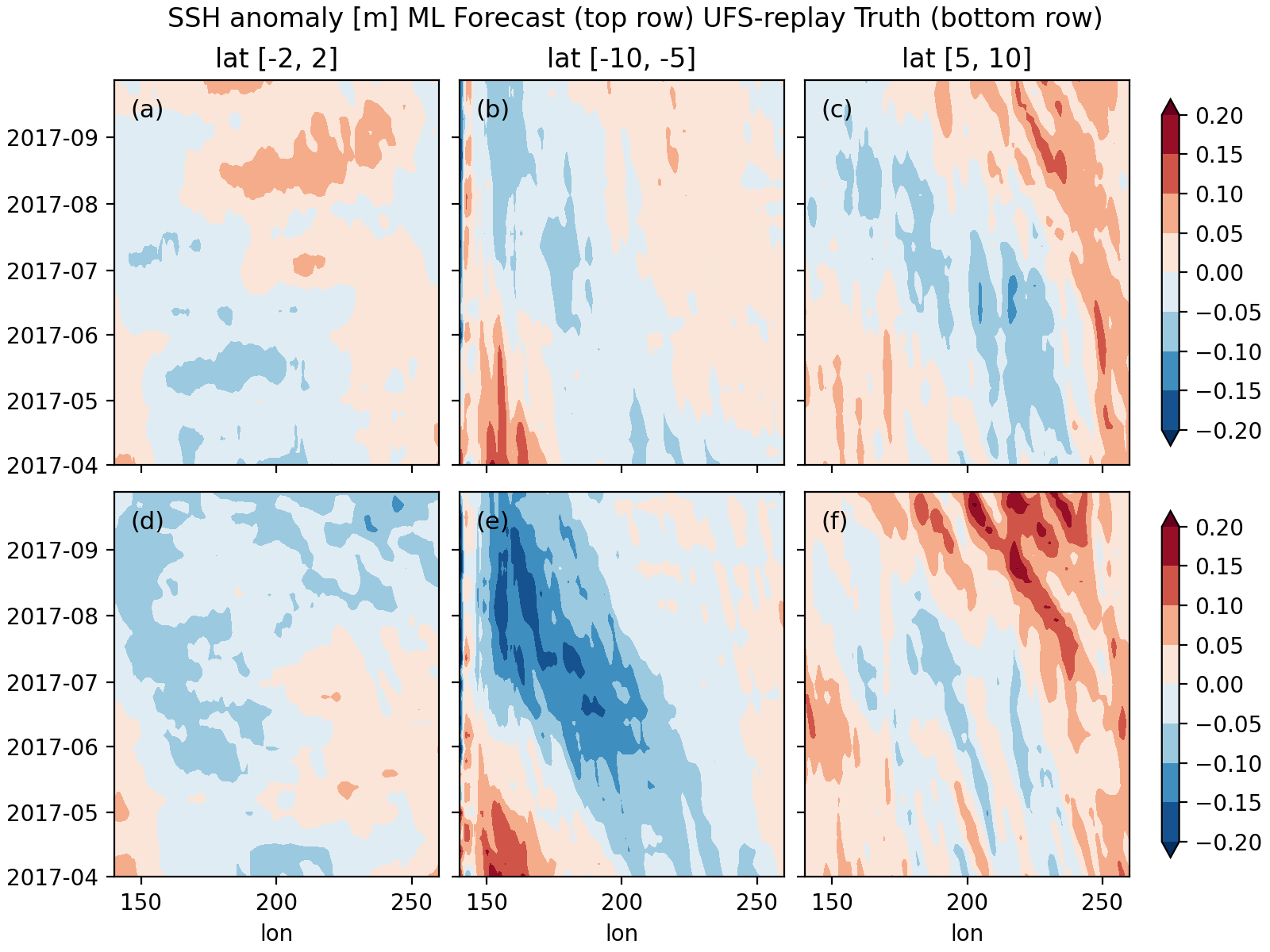}
    \caption{Same as Figure~\ref{fig:eqwave} but for forecast initialized at 2017/04/01-00UTC}
    \label{fig:app_ocn_wave_201704}
\end{figure}
\begin{figure}
    \centering
    \includegraphics[width=0.7\textwidth]{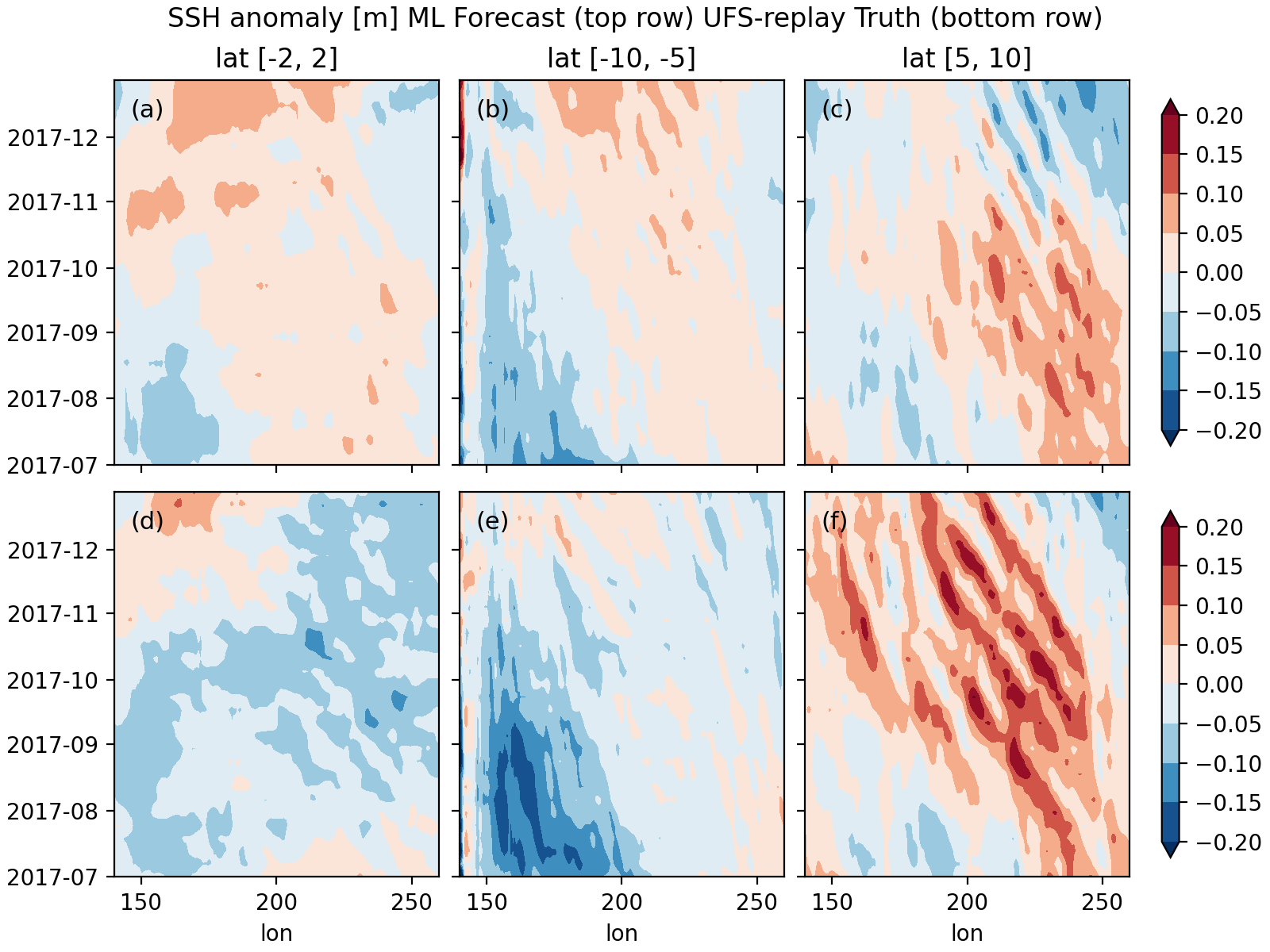}
    \caption{Same as Figure~\ref{fig:eqwave} but for forecast initialized at 2017/07/01-00UTC}
    \label{fig:app_ocn_wave_201707}
\end{figure}
\begin{figure}
    \centering
    \includegraphics[width=0.7\textwidth]{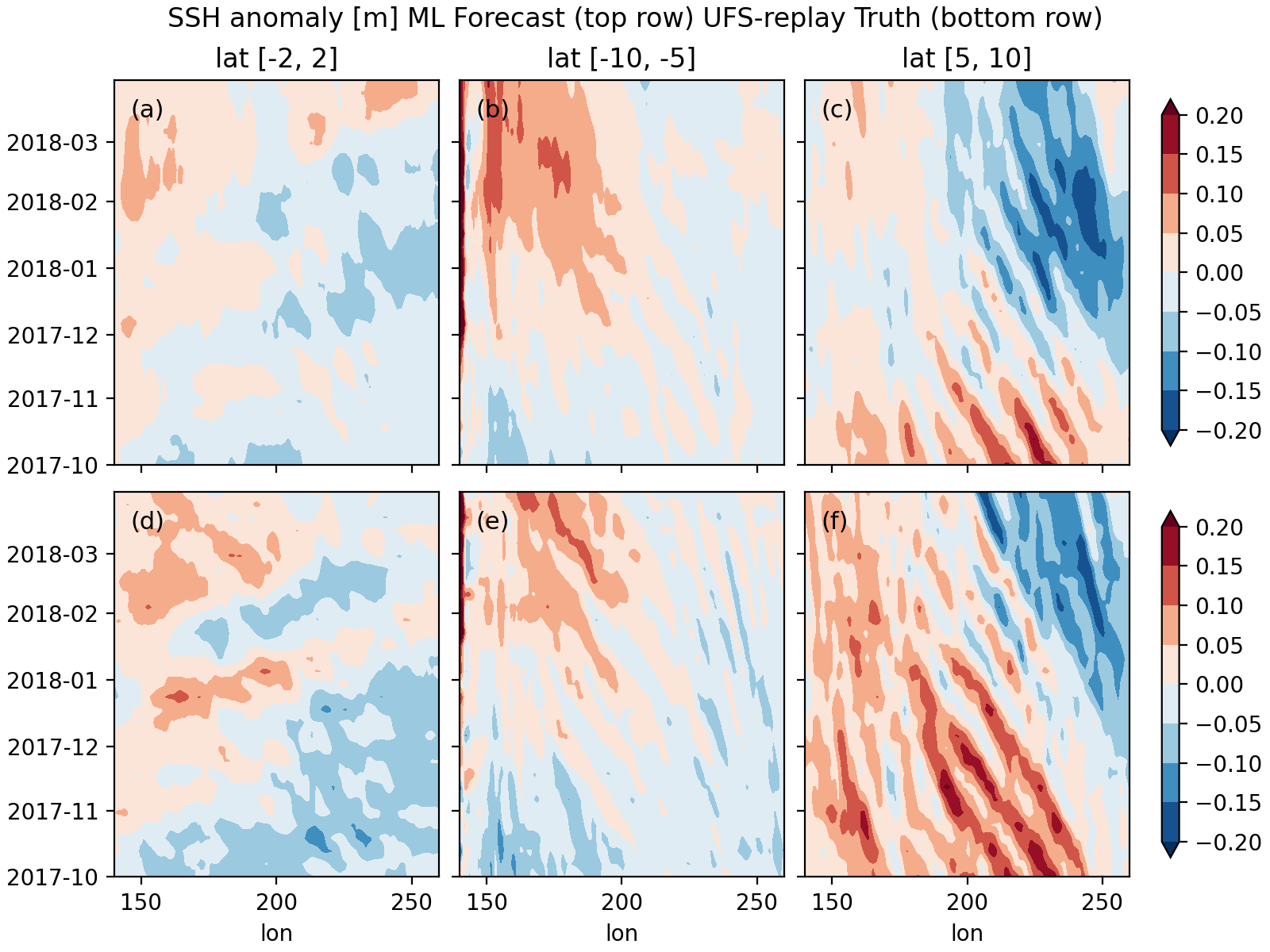}
    \caption{Same as Figure~\ref{fig:eqwave} but for forecast initialized at 2017/10/01-00UTC}
    \label{fig:app_ocn_wave_201710}
\end{figure}

\begin{figure}
    \centering
    \includegraphics[width=0.7\textwidth]{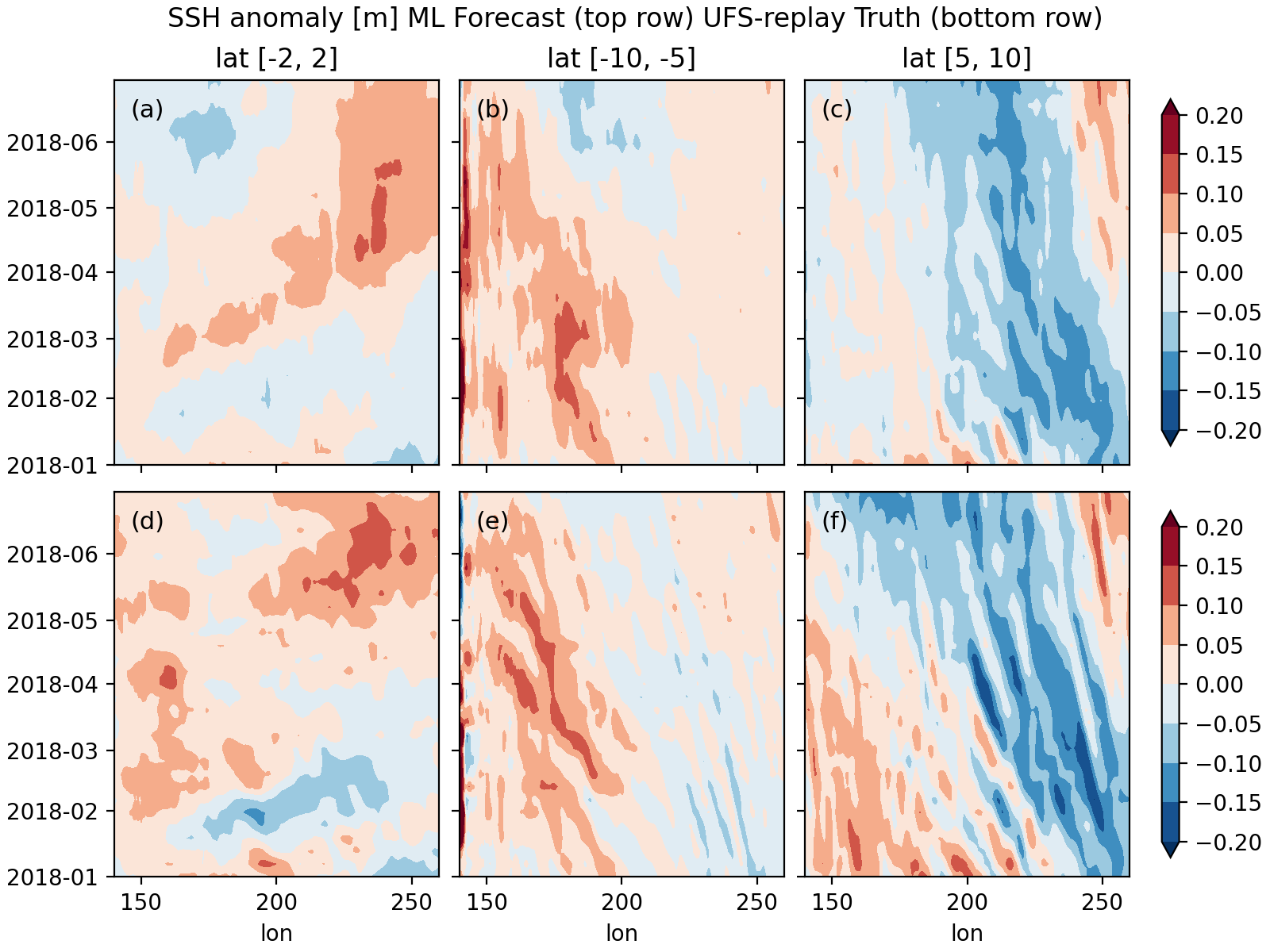}
    \caption{Same as Figure~\ref{fig:eqwave} but for forecast initialized at 2018/01/01-00UTC}
    \label{fig:app_ocn_wave_201801}
\end{figure}
\begin{figure}
    \centering
    \includegraphics[width=0.7\textwidth]{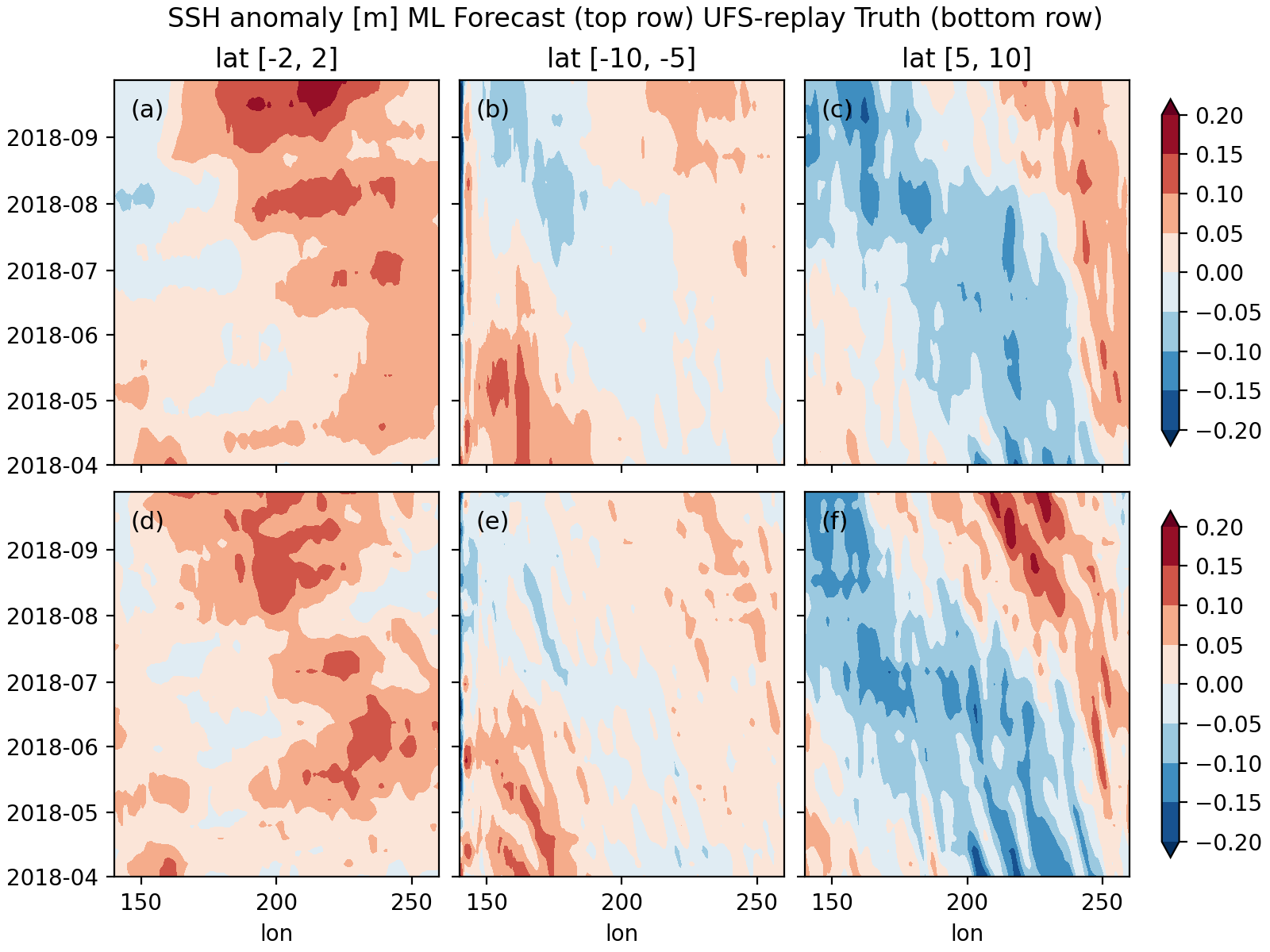}
    \caption{Same as Figure~\ref{fig:eqwave} but for forecast initialized at 2018/04/01-00UTC}
    \label{fig:app_ocn_wave_201804}
\end{figure}
\begin{figure}
    \centering
    \includegraphics[width=0.7\textwidth]{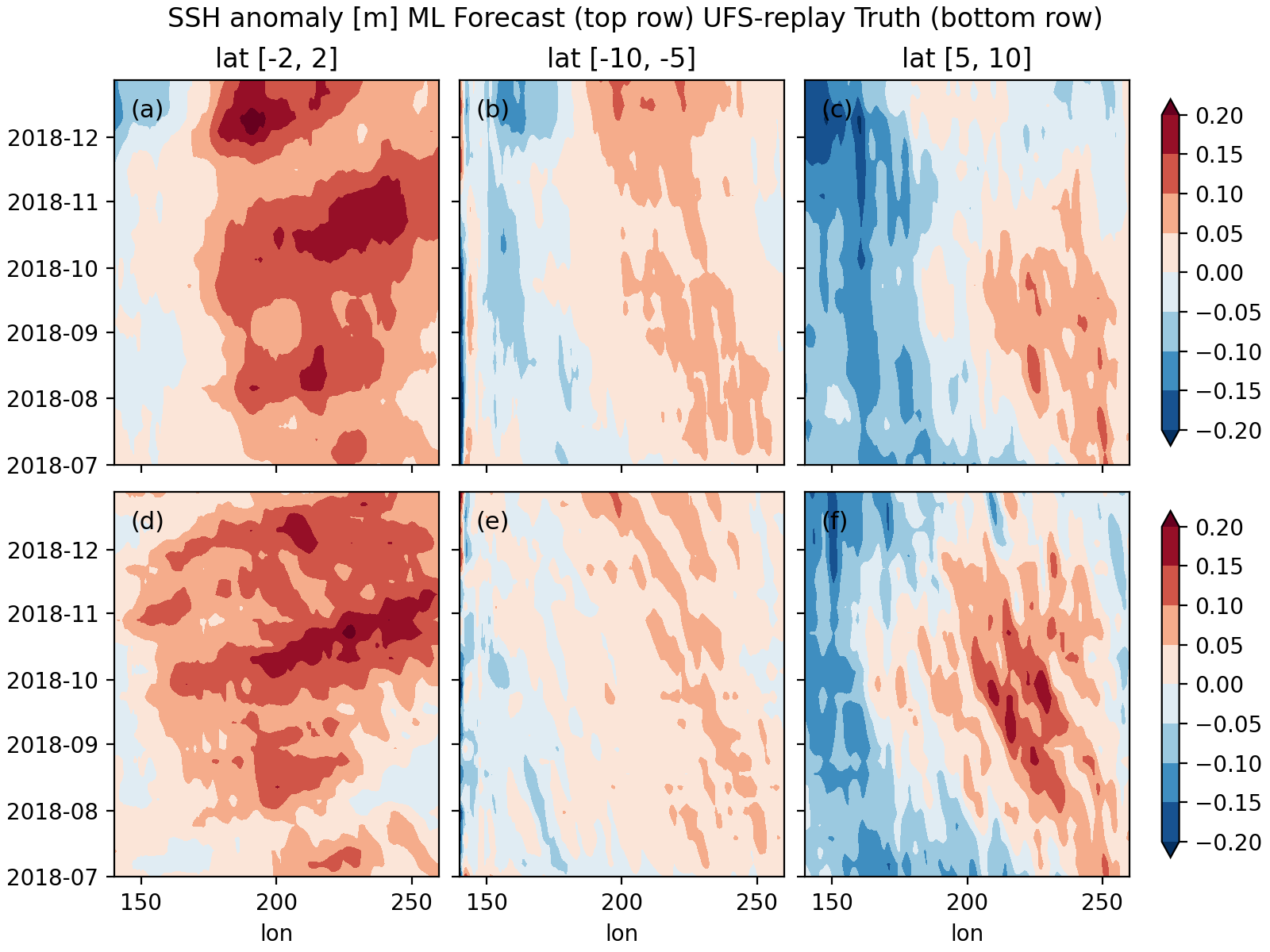}
    \caption{Same as Figure~\ref{fig:eqwave} but for forecast initialized at 2018/07/01-00UTC}
    \label{fig:app_ocn_wave_201807}
\end{figure}
\begin{figure}
    \centering
    \includegraphics[width=0.7\textwidth]{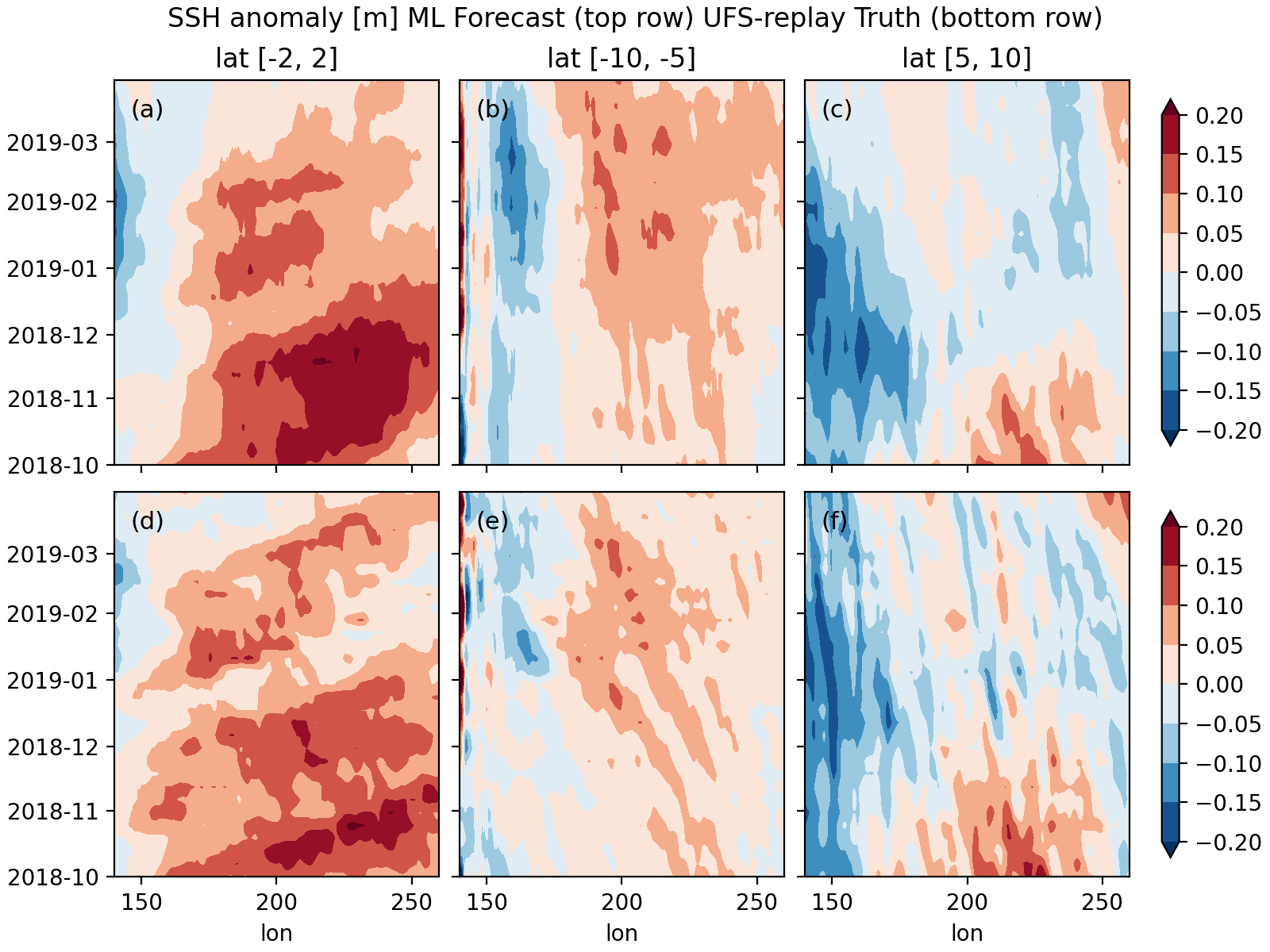}
    \caption{Same as Figure~\ref{fig:eqwave} but for forecast initialized at 2018/10/01-00UTC}
    \label{fig:app_ocn_wave_201810}
\end{figure}

\begin{figure}
    \centering
    \includegraphics[width=0.7\textwidth]{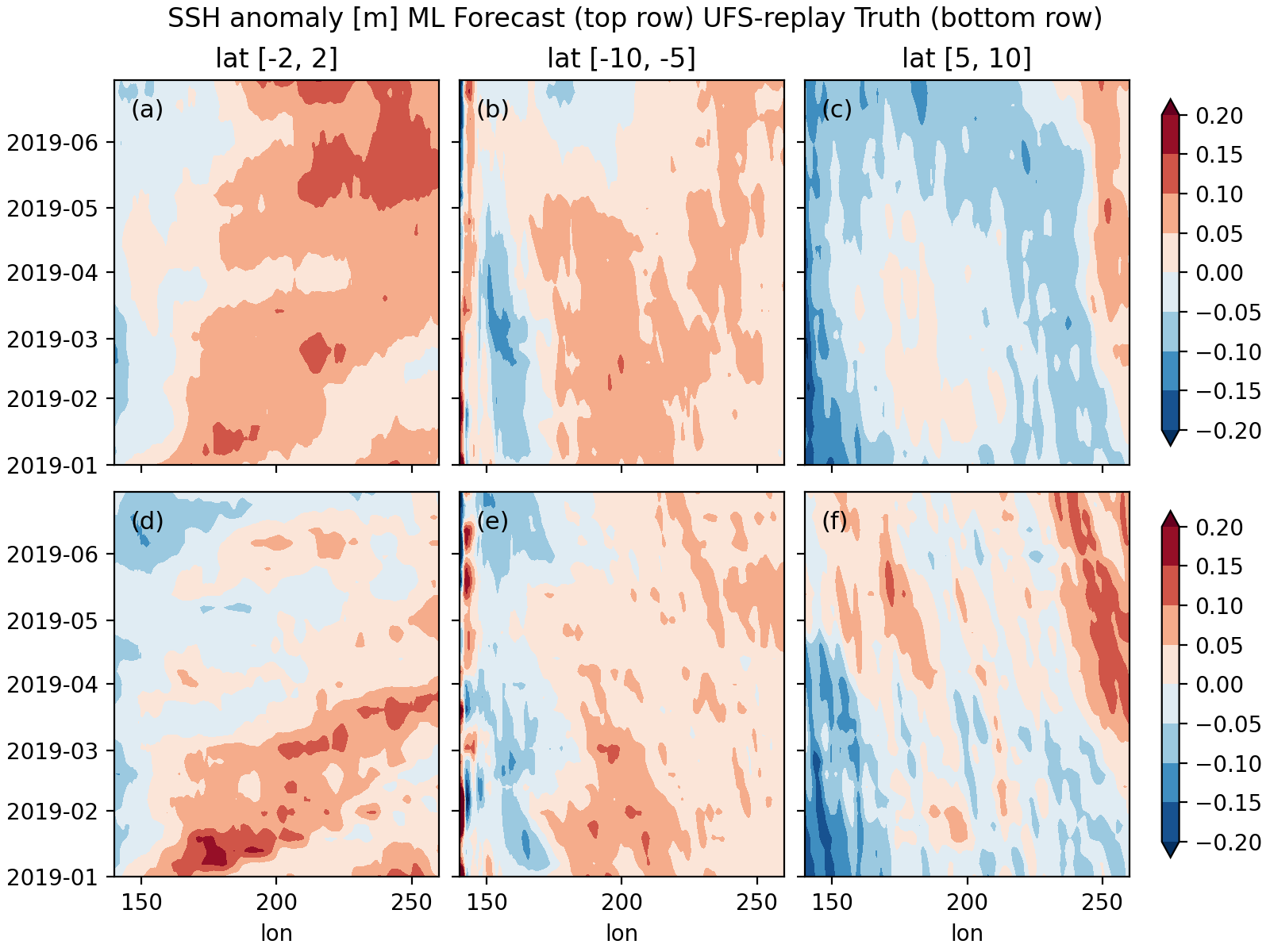}
    \caption{Same as Figure~\ref{fig:eqwave} but for forecast initialized at 2019/01/01-00UTC}
    \label{fig:app_ocn_wave_201901}
\end{figure}
\begin{figure}
    \centering
    \includegraphics[width=0.7\textwidth]{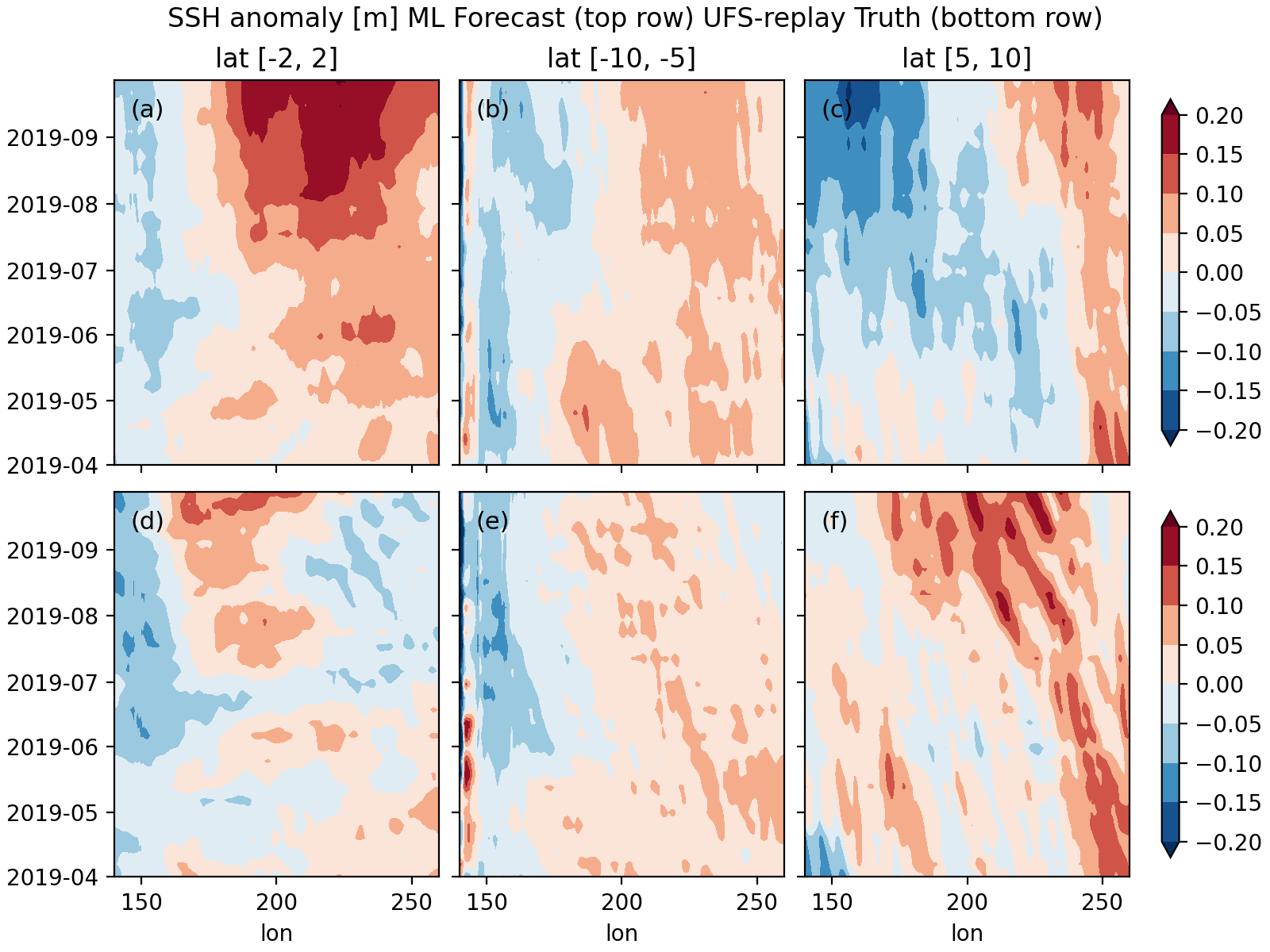}
    \caption{Same as Figure~\ref{fig:eqwave} but for forecast initialized at 2019/04/01-00UTC}
    \label{fig:app_ocn_wave_201904}
\end{figure}
\begin{figure}
    \centering
    \includegraphics[width=0.7\textwidth]{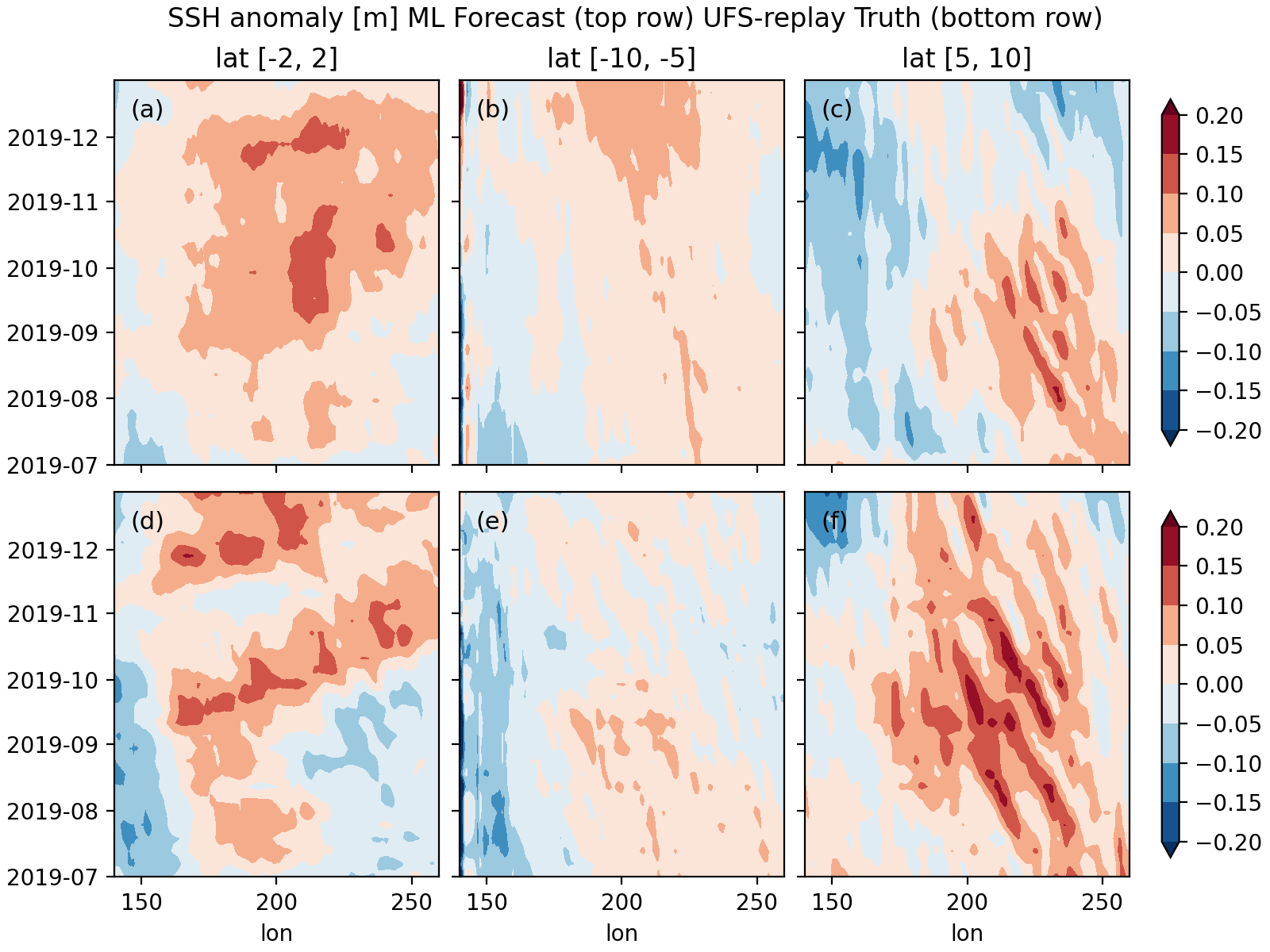}
    \caption{Same as Figure~\ref{fig:eqwave} but for forecast initialized at 2019/07/01-00UTC}
    \label{fig:app_ocn_wave_201907}
\end{figure}
\begin{figure}
    \centering
    \includegraphics[width=0.7\textwidth]{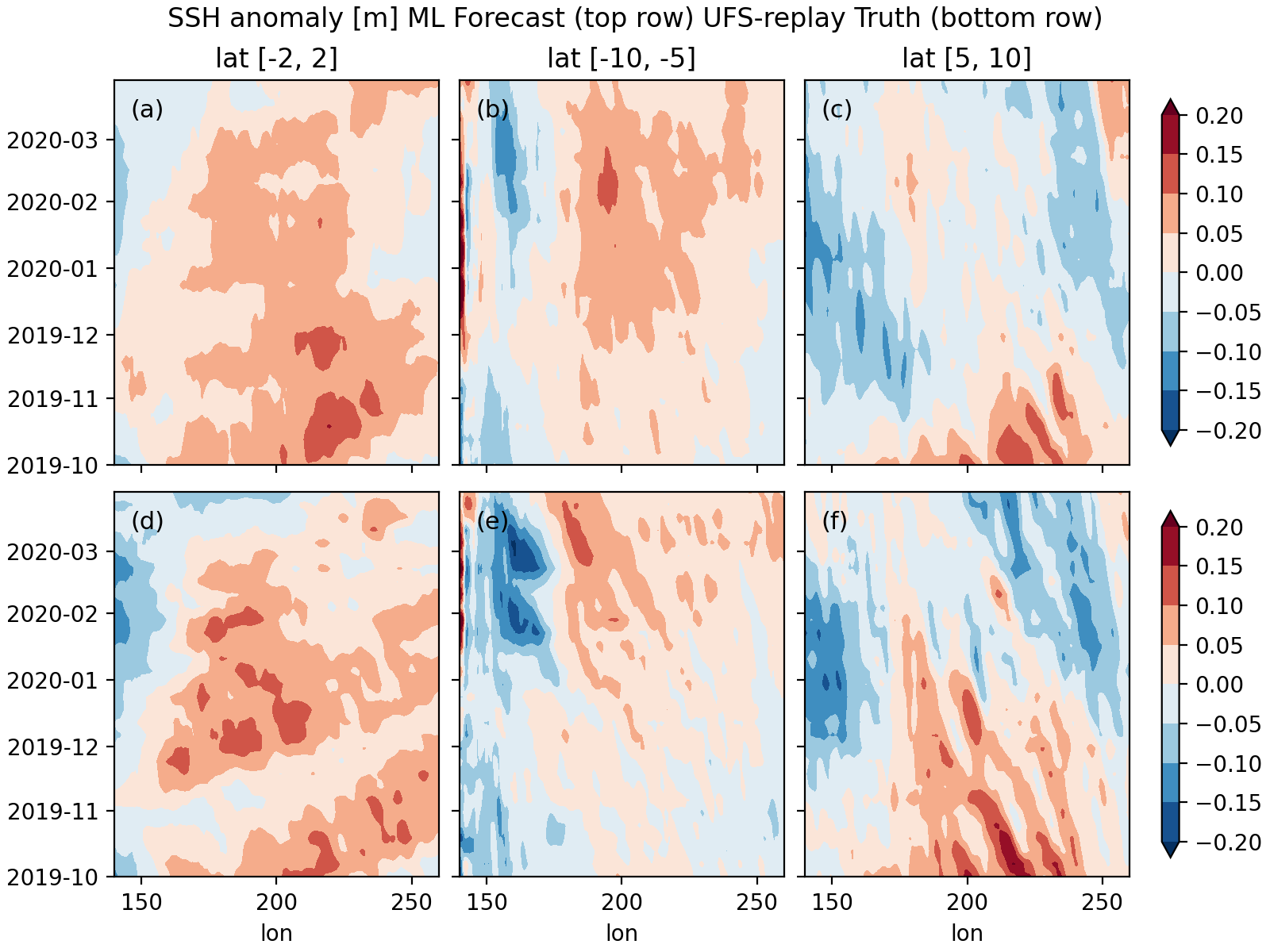}
    \caption{Same as Figure~\ref{fig:eqwave} but for forecast initialized at 2019/10/01-00UTC}
    \label{fig:app_ocn_wave_201910}
\end{figure}

\clearpage
\refstepcounter{AppendixCounter} 
\section*{Appendix \theAppendixCounter}\label{sec:app_bias}
\setcounter{figure}{0}
Lead-time dependant model biases are computed using the monthly hindcast initialized from 1994 to 2016, the same period used to train the model. The same LEF method is used to generate 3312 forecasts in total. Then, we compute model biases by subtracting the ensemble mean and initialization mean of the simulated results from the truth (training data) on lead-time and spatial dimensions for all variables. When applying the bias correction on the out-of-sample years (2017-2021), we remove the pre-computed biases from the results.

\begin{figure}[h]
    \centering
    \includegraphics[width=0.7\textwidth]{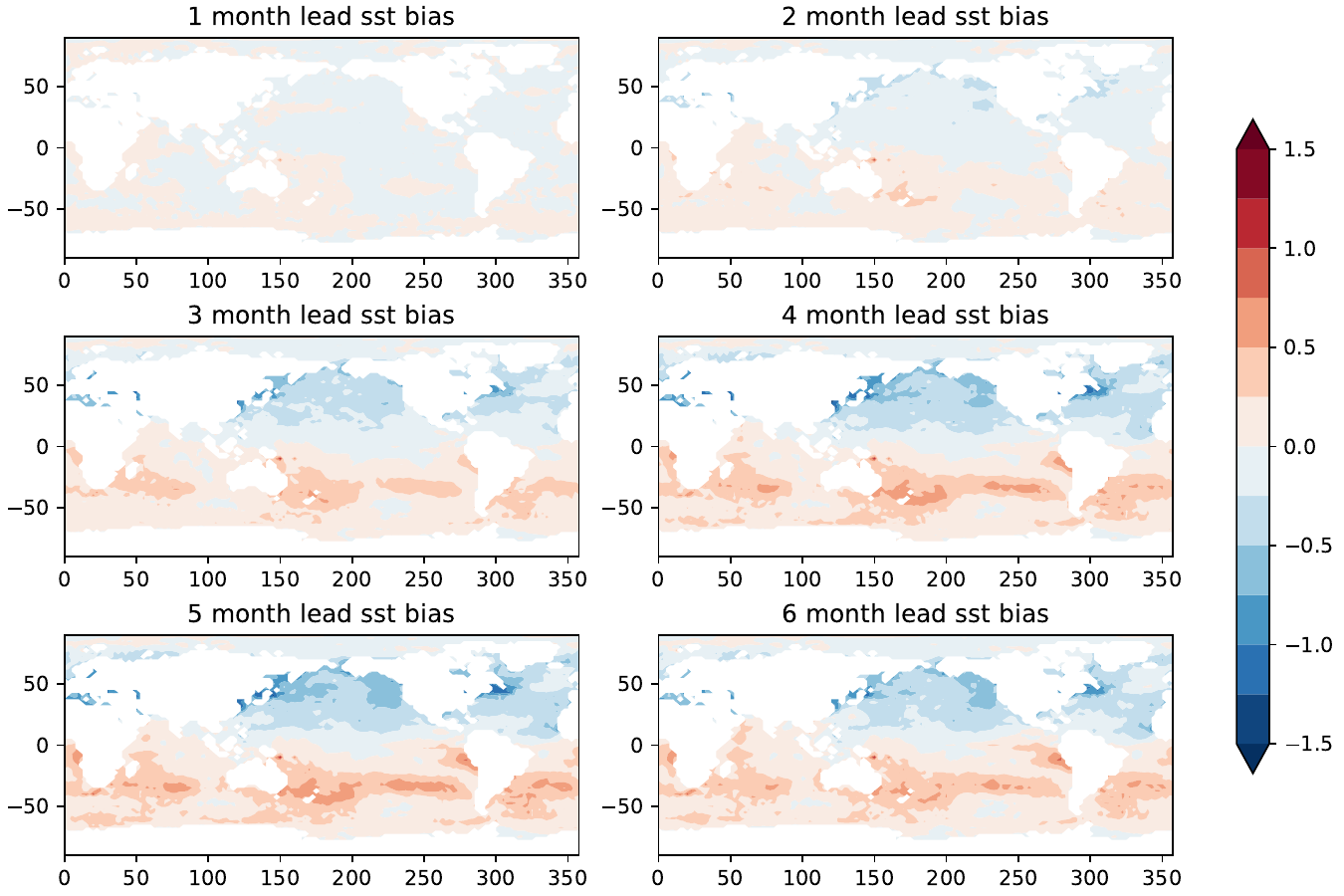}
    \caption{Lead time bias of SST [K].}
    \label{fig:app_bias_sst}
\end{figure}
\begin{figure}
    \centering
    \includegraphics[width=0.7\textwidth]{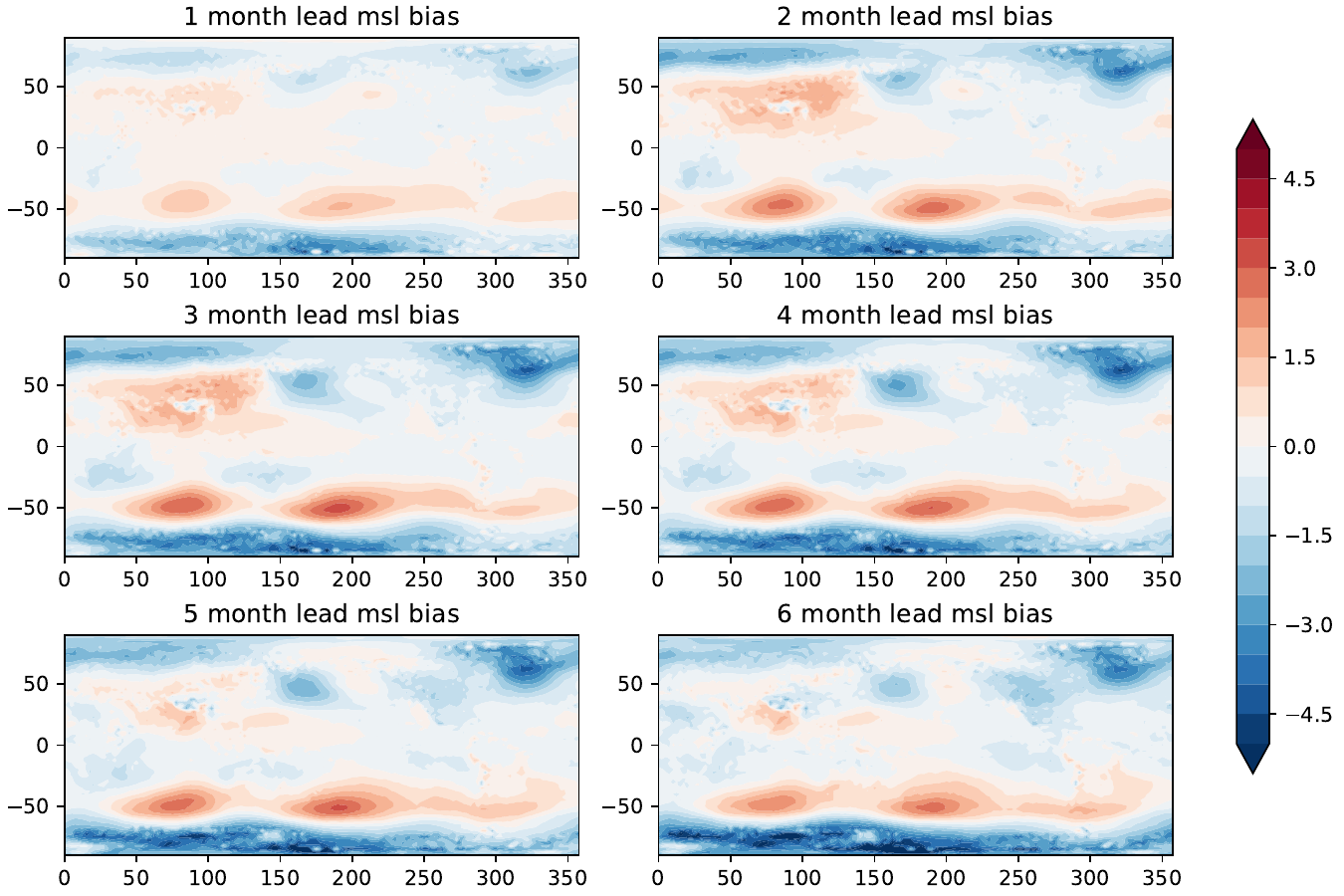}
    \caption{Lead time bias of MSLP[hPa].}
    \label{fig:app_bias_mslp}
\end{figure}
\begin{figure}
    \centering
    \includegraphics[width=0.7\textwidth]{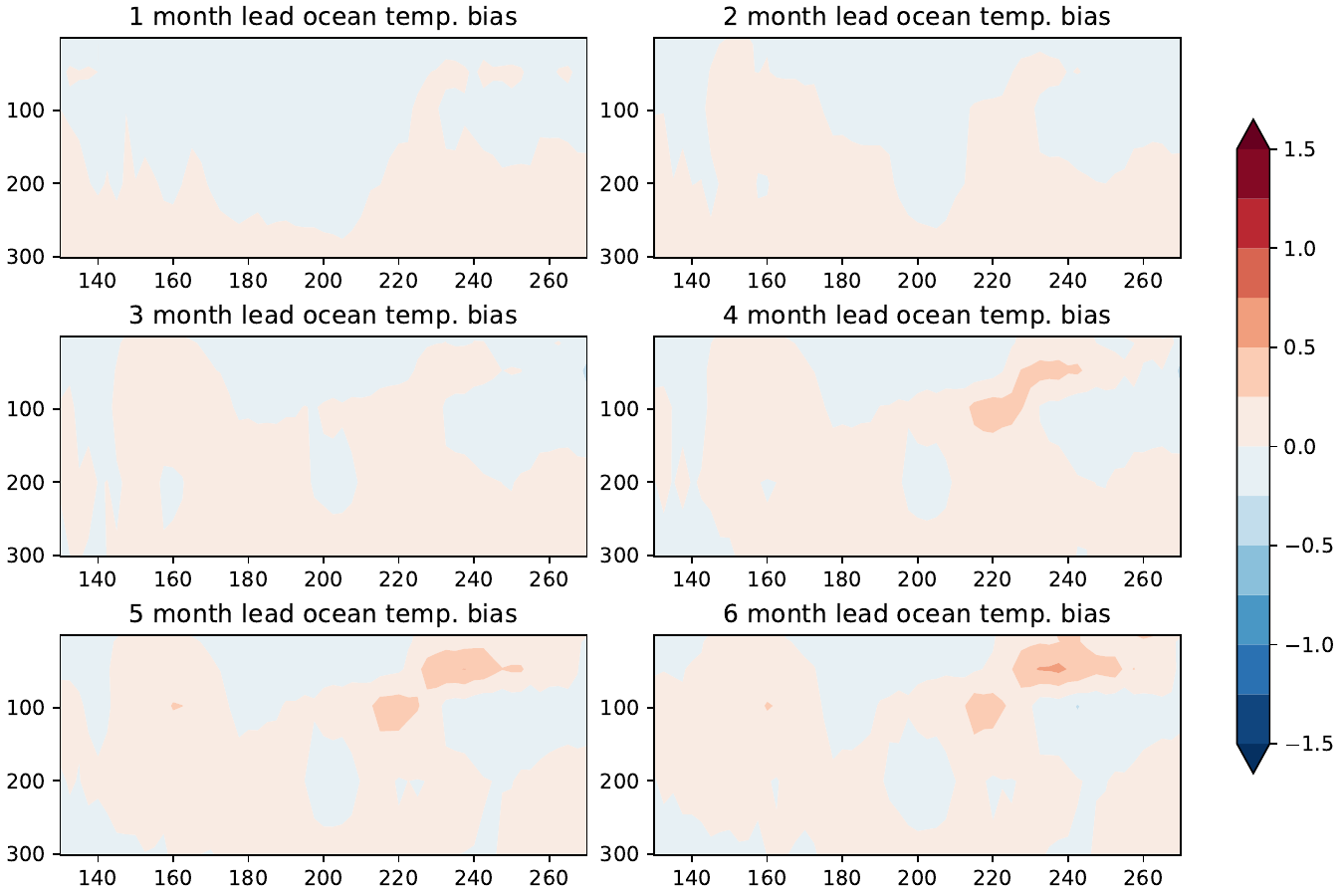}
    \caption{Lead time bias of equatorial Pacific upper ocean temperature [K].}
    \label{fig:app_bias_to}
\end{figure}

\clearpage
\refstepcounter{AppendixCounter} 
\section*{Appendix \theAppendixCounter}\label{sec:app_atmospherecomposites}
\setcounter{figure}{0}
\textbf{Ocean El Niño and La Niña composites}
\begin{figure}[h]
    \centering
    \includegraphics[width=0.7\textwidth]{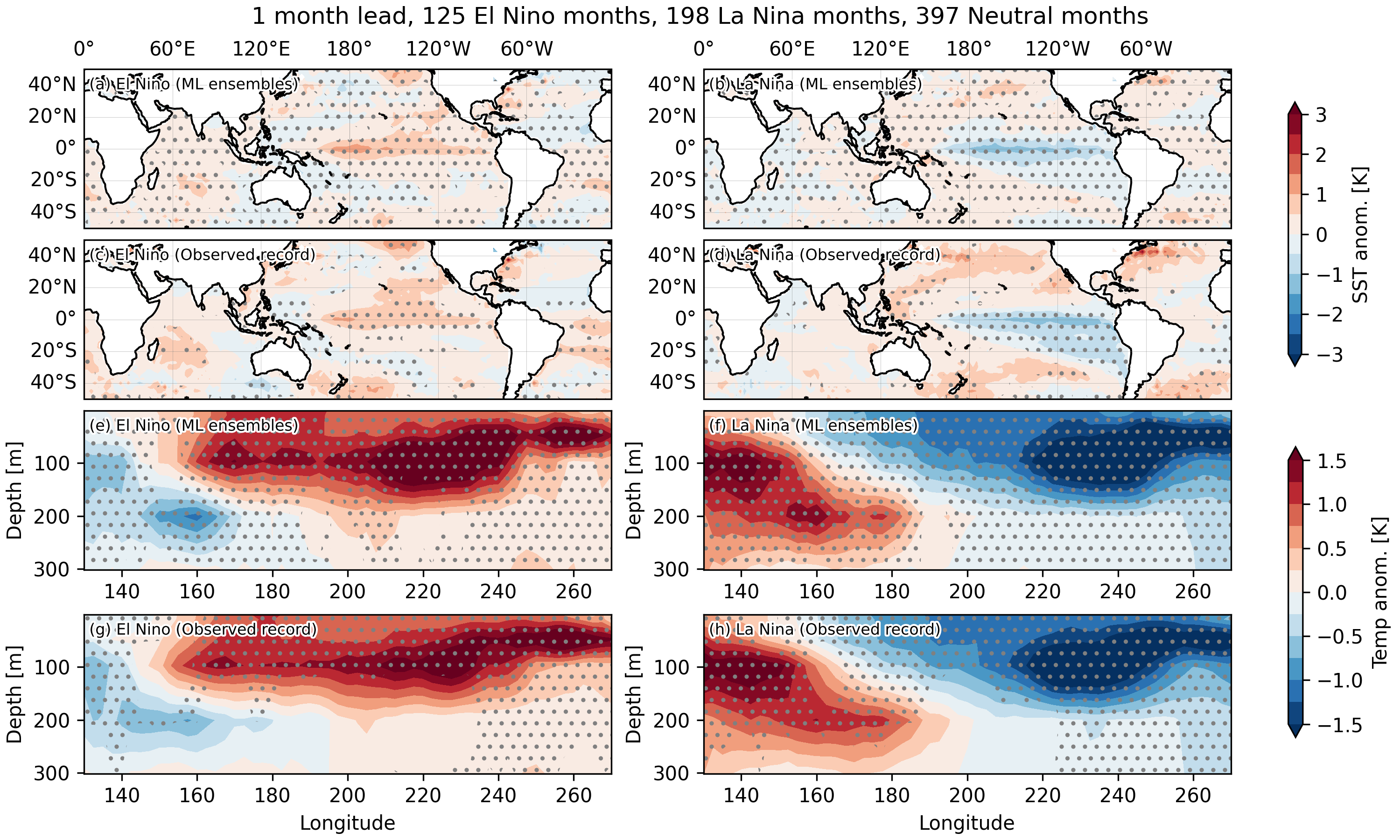}
    \caption{Forecast lead time: 1 month}
    \label{fig:app_ocn_1m}
\end{figure}

\begin{figure}
    \centering
    \includegraphics[width=0.7\textwidth]{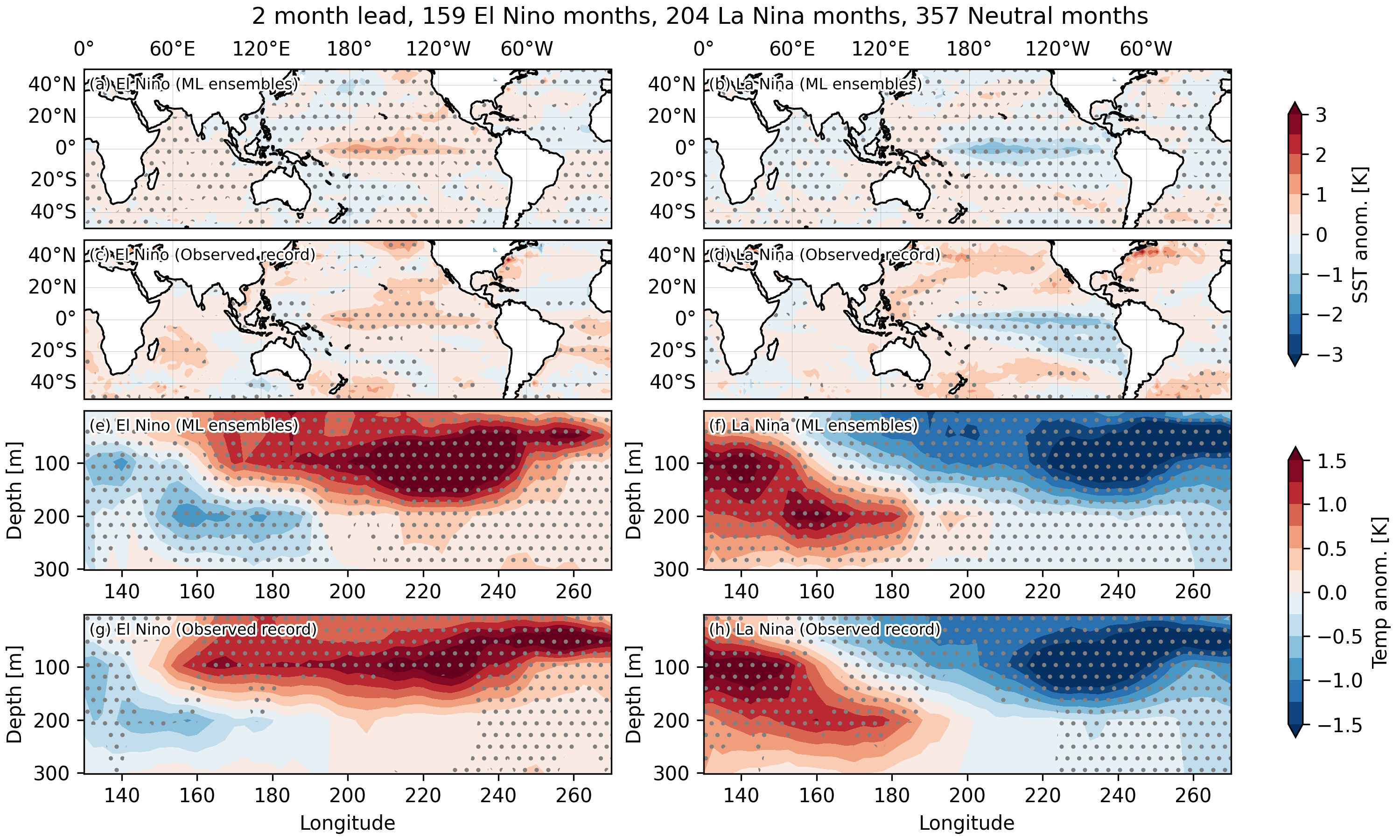}
    \caption{Forecast lead time: 2 months}
    \label{fig:app_ocn_2m}
\end{figure}

\begin{figure}
    \centering
    \includegraphics[width=0.7\textwidth]{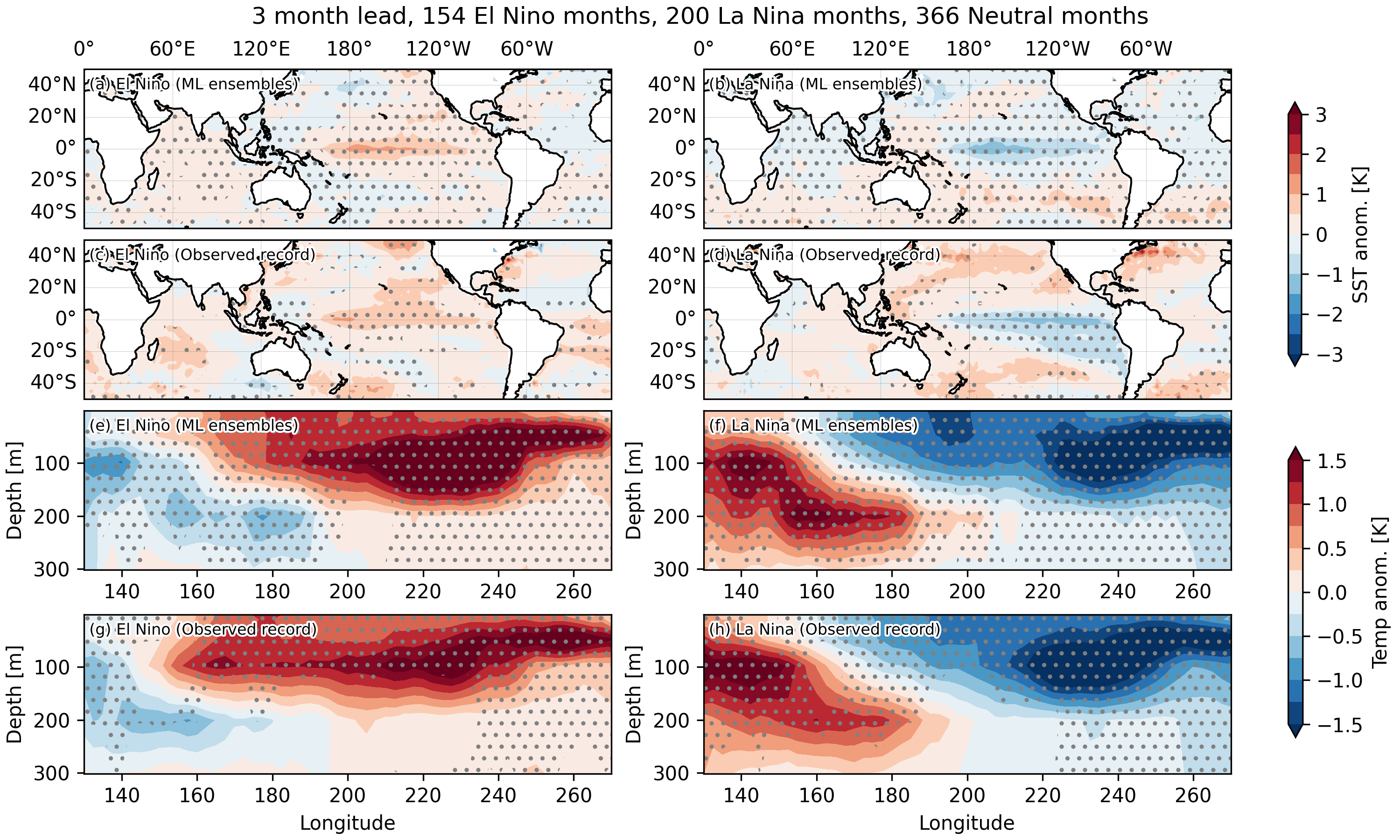}
    \caption{Forecast lead time: 3 months}
    \label{fig:app_ocn_3m}
\end{figure}

\begin{figure}
    \centering
    \includegraphics[width=0.7\textwidth]{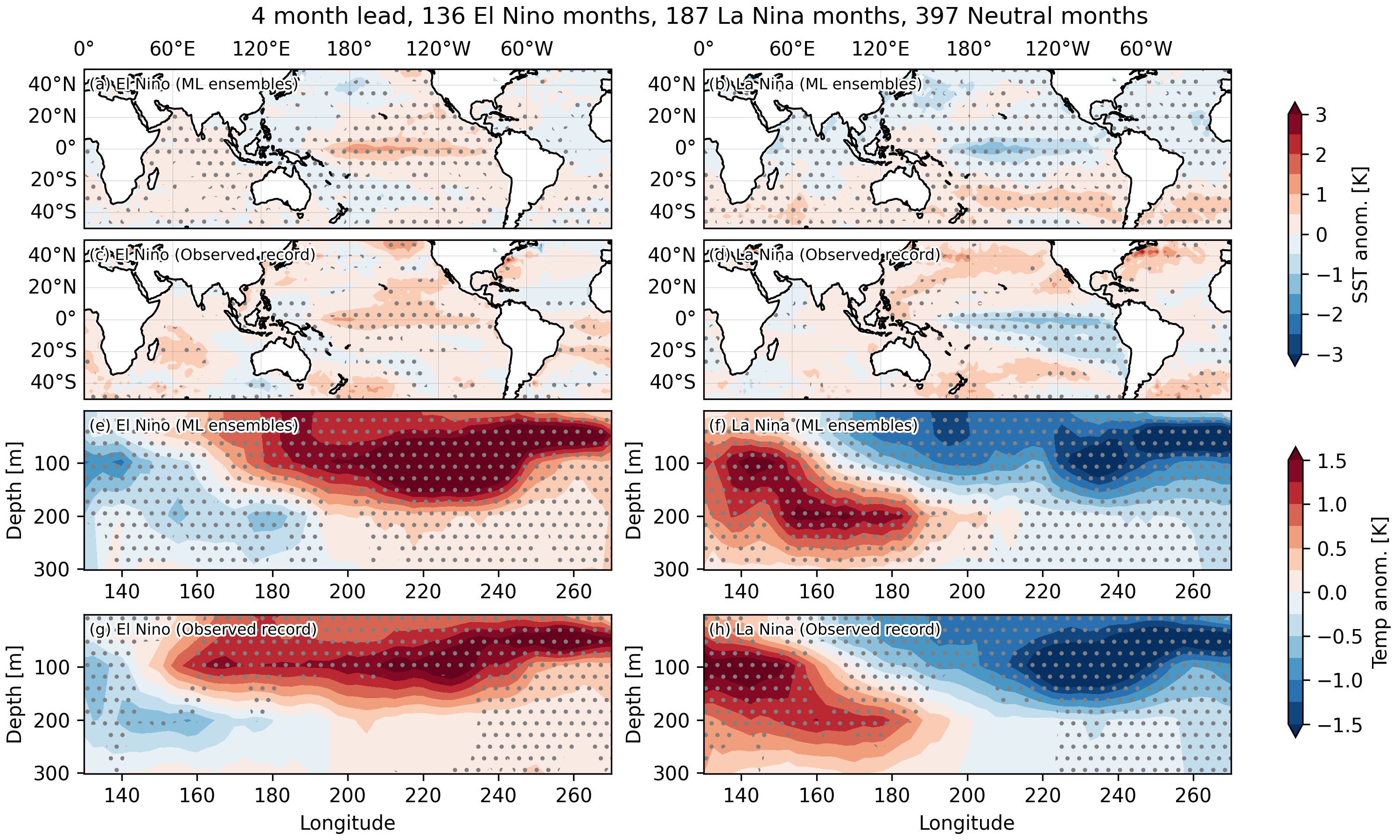}
    \caption{Forecast lead time: 4 months}
    \label{fig:app_ocn_4m}
\end{figure}

\begin{figure}
    \centering
    \includegraphics[width=0.7\textwidth]{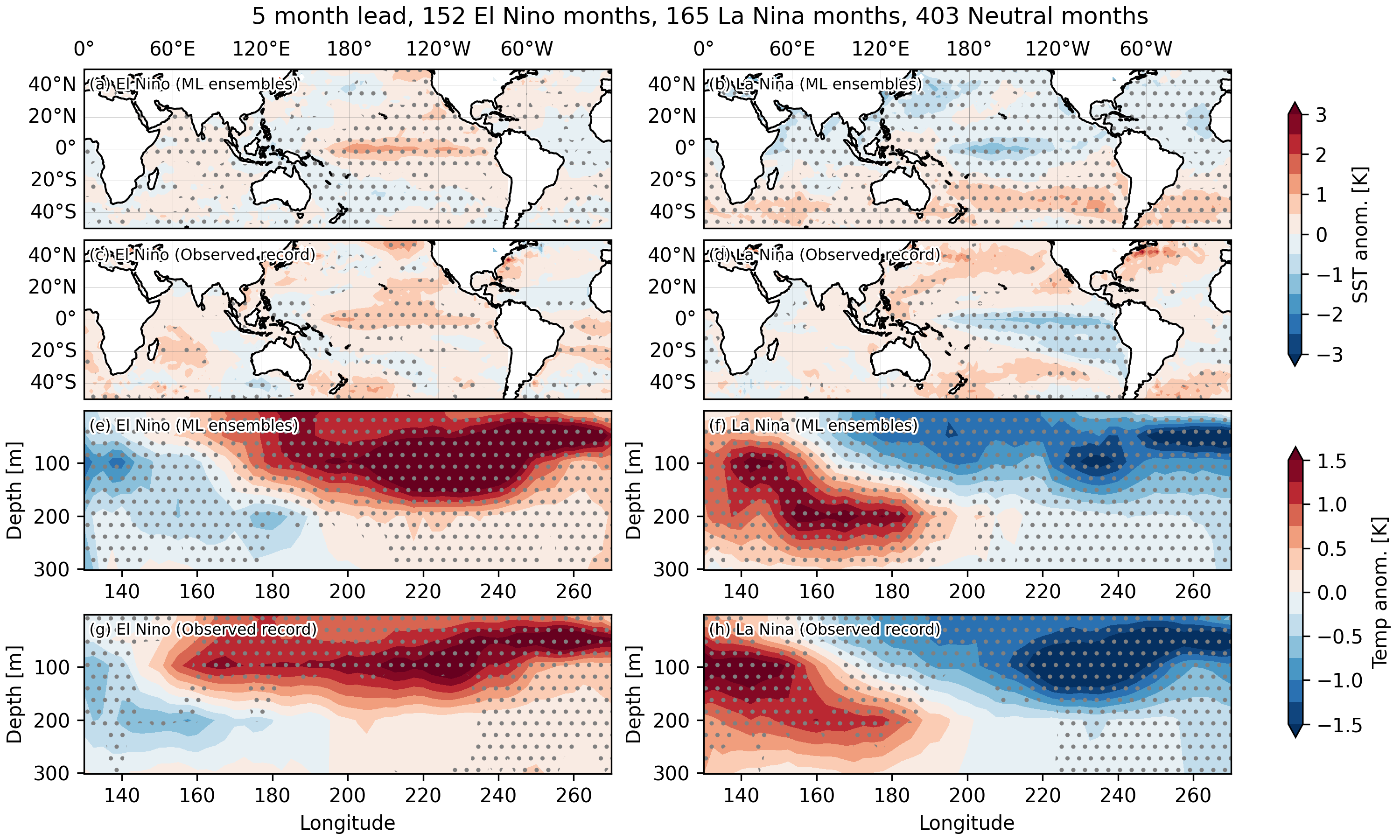}
    \caption{Forecast lead time: 5 months}
    \label{fig:app_ocn_5m}
\end{figure}

\begin{figure}
    \centering
    \includegraphics[width=0.7\textwidth]{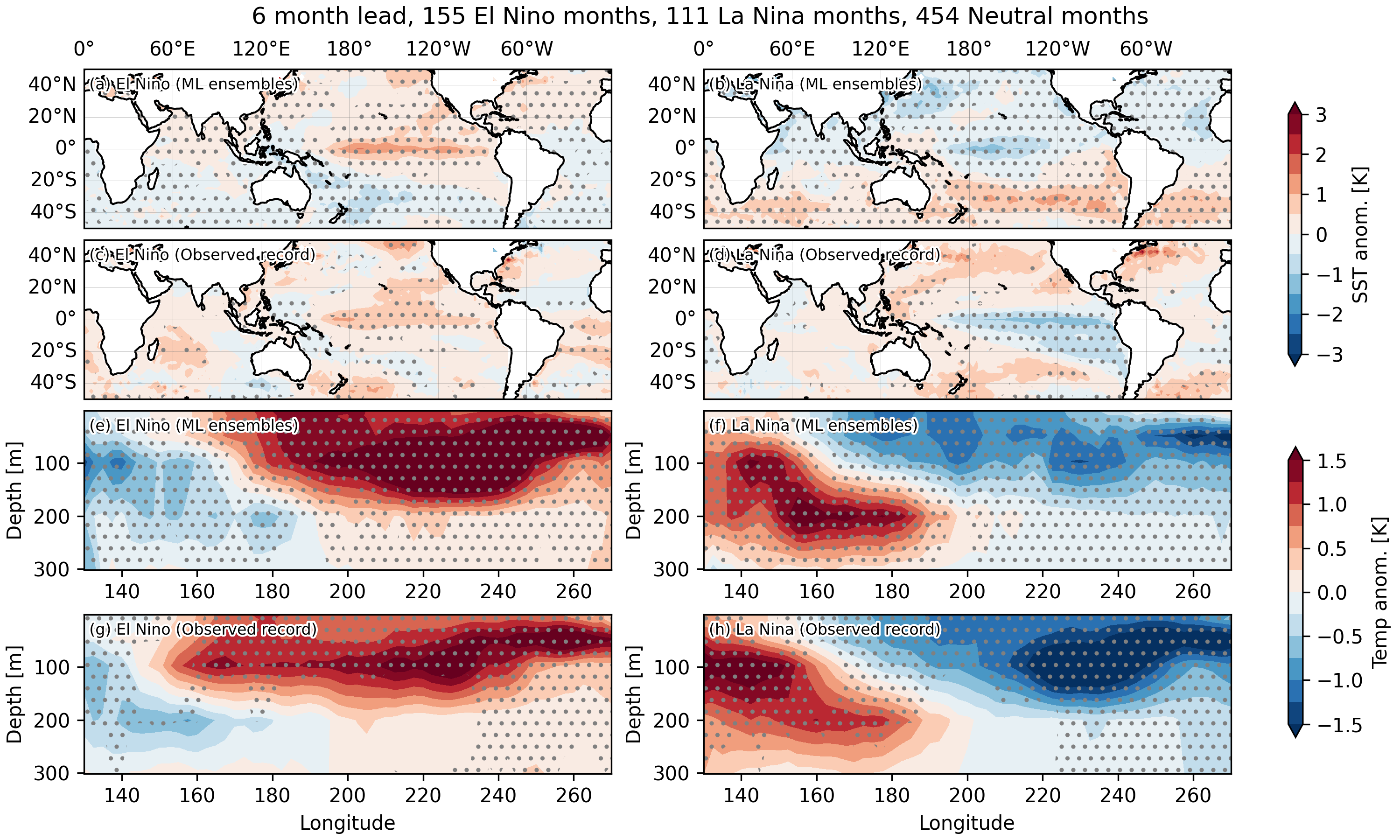}
    \caption{Forecast lead time: 6 months}
    \label{fig:app_ocn_6m}
\end{figure}
 
\end{document}